\documentclass[12pt,oneside, a4paper]{article}

\ifx\pdfoutput\undefined
\usepackage[dvips,bookmarks=false]{hyperref}	
\else
\usepackage{hyperref}	
\fi
\hypersetup{colorlinks,bookmarksopen,bookmarksnumbered,citecolor=blue,
linkcolor=black,pdfstartview=FitH,urlcolor=blue}

\usepackage{graphicx}
\usepackage{amssymb}
\usepackage{cite}
\usepackage{bm}
\usepackage{indentfirst}
\usepackage{amsmath}
\usepackage{hhline}
\usepackage{multirow}

\usepackage{ulem}

\begin{document}

\begin{titlepage}

\begin{flushright}
KANAZAWA-16-12\\
LPT-Orsay-16-62\\
TUM-HEP-1070-16
\end{flushright}

\begin{center}

\vspace{1cm}
{\large\bf 
Implications of Two-component Dark Matter Induced by Forbidden Channels
 and Thermal Freeze-out}
\vspace{1cm}

\renewcommand{\thefootnote}{\fnsymbol{footnote}}
Mayumi Aoki$^1$\footnote[1]{mayumi@hep.s.kanazawa-u.ac.jp}
,
Takashi Toma$^{2,3}$\footnote[2]{takashi.toma@tum.de}
\vspace{5mm}

{\it%
$^1${Institute for Theoretical Physics, Kanazawa University, Kanazawa
 920-1192, Japan}
$^2${Laboratoire de Physique Th\'eorique, CNRS,\\ 
Univ. Paris-Sud, Universit\'e Paris-Saclay, 91405 Orsay, France}\\
$^3${Physik-Department T30d, Technische Universit\"at M\"unchen,\\
 James-Franck-Stra\ss{}e, D-85748 Garching, Germany}
}

\vspace{8mm}

\abstract{
We consider a model of two-component dark matter based on a hidden
 $U(1)_D$ symmetry, in which relic densities of the dark matter are determined by forbidden
 channels and thermal freeze-out. 
The hidden $U(1)_D$ symmetry is spontaneously broken to a residual
 $\mathbb{Z}_4$ symmetry, and the lightest $\mathbb{Z}_4$ charged
 particle can be a dark matter candidate. 
Moreover, depending on the mass hierarchy in the dark sector, we
 have two-component dark matter. 
We show that the relic density of the lighter dark matter component
can be determined by forbidden
 annihilation channels which require larger couplings compared to the
 normal freeze-out mechanism. 
As a result, a large self-interaction of the lighter dark matter
 component can be induced, which may solve small scale problems of $\Lambda$CDM model. 
 On the other hand, the heavier dark matter component is produced by normal freeze-out
 mechanism. We find that interesting implications emerge between the two dark matter
 components in this framework. 
We explore detectabilities of these dark matter particles and show some parameter space
can be tested by the SHiP experiment. 
 }

\end{center}
\end{titlepage}

\renewcommand{\thefootnote}{\arabic{footnote}}
\newcommand{\bhline}[1]{\noalign{\hrule height #1}}
\newcommand{\bvline}[1]{\vrule width #1}

\setcounter{footnote}{0}

\setcounter{page}{1}

\section{Introduction}
There are some clear evidences for the existence of dark matter in the
universe such as the rotation curves of spiral galaxies, the gravitational lensing
effects, the cosmic microwave background (CMB) measurements, the large scale
structure of the universe and the collision of the bullet clusters. 
From the view point of particle physics, the most well-known and
promising dark matter candidate would be the Weakly 
Interacting Massive Particles (WIMPs) with the mass of electroweak to TeV scale. 
This kind of dark matter has been explored through direct,
indirect and collider searches. However in spite of making a great effort
to find WIMPs, any clear signal of WIMPs is not found yet.

Self-interactions of dark matter which are different from gravitational force may be a
key to understand the nature of dark matter. 
Self-interactions may give a solution for the small scale problems of 
the standard 
cosmological model (the so-called $\Lambda$CDM model)
such as
the cusp-core, too-big-to-fail and missing satellites
problems. 
The cusp-core problem is that the profile of dark matter 
inferred by N-body simulation with collisionless dark matter does not
match with that estimated from 
the observation of the rotation curves of the spiral galaxies. 
The too-big-to-fail problem is that subhalos inferred by simulations with cold
dark matter is too dense compared to the observations in the Milky Way. 
The missing satellites problem is that the predicted number of the subhalos in the local group 
is an order of magnitude more  than the  number of the observed satellites.
The required magnitude of the self-interacting cross section in order to solve
these problems is 
roughly given by $\sigma/m\sim0.5-50~\mathrm{cm^2/g}$ where $\sigma$ is
the self-interacting cross section and $m$ is the dark matter
mass~\cite{Elbert:2014bma}.
One should note that the collision of the bullet
clusters gives an upper bound on the self-interacting cross section as
$\sigma/m<1.25~\mathrm{cm^2/g}$~\cite{Randall:2007ph}. 
Another implication of self-interacting dark matter is the recent
observation of the cluster Abell 3827 at which it is measured an off-set of
$1.62~\mathrm{kpc}$ between the center of the dark 
matter sub-halo and the galaxy with $3.3~\sigma$ confidence
level~\cite{Massey:2015dkw}. 
This may be understood with a large self-interaction of dark matter, and 
the required magnitude of the self-interacting cross section is
$\sigma/m\sim\mathcal{O}(1)~\mathrm{cm^2/g}$~\cite{Massey:2015dkw, 
Kahlhoefer:2015vua}. 

Understanding such a large self-interaction of dark matter with typical
WIMPs may be difficult since the ratio of cross section and dark 
matter mass $\sigma/m$ is roughly scaled by $\sigma/m\propto m^{-3}$, thus the
ratio $\sigma/m$ sharply decreases with increasing dark mater mass. 
A Strongly Interacting Massive Particle (SIMP) is one of the good
candidates which can have large self-interacting cross
section~\cite{Hochberg:2014dra, Hochberg:2014kqa, Bernal:2015ova,
Bernal:2015bla, Choi:2015bya, Bernal:2015xba, Hochberg:2015vrg,
Choi:2016hid, Choi:2016tkj}.\footnote{Introducing $\mathbb{Z}_3$ symmetry is the
simplest extension of SIMP dark matter. 
Getting a large self-interacting cross section for the SIMP 
in a $\mathbb{Z}_3$ symmetric model, however, may not be easy consistently with
perturbative couplings and potential stability~\cite{Choi:2015bya}. 
} 
In addition, the candidate which is a kind of WIMPless dark matter~\cite{Feng:2008ya}
produced by forbidden channels is another possibility to obtain a large
self-interacting cross section~\cite{Choi:2016tkj, D'Agnolo:2015koa, Delgado:2016umt}.

On the other hand, although the dark matter existing in the universe is normally
assumed to be occupied by one-component for
simplicity, this assumption is not necessary and the dark matter can be
composed of multi-particles in general. 
If multi-component dark matter is assumed, some interesting
phenomenology is expected to occur. 
For example, if the second dark matter component is sub-dominant
and has large self-interaction, this sub-dominant
component may form a disk like the normal
matter, which
is discussed as Double Disk Dark Matter in Refs.~\cite{Fan:2013yva,
Fan:2013tia}. In this case, different properties for indirect and direct
detection of dark matter would emerge. 
Other implications of multi-component dark matter also have been
explored~\cite{Dienes:2014via}. 
Some UV complete models with multi-component dark matter and its
phenomenology have been
explored based on the simplest $\mathbb{Z}_4$
symmetry~\cite{Cai:2015zza}, radiative neutrino 
masses~\cite{Aoki:2014lha, Aoki:2013gzs, Kajiyama:2013rla, Wang:2015saa, 
Ho:2016aye}, hidden gauge symmetries~\cite{Foot:2014uba, Gross:2015cwa,
Arcadi:2016kmk}, $\mathbb{Z}_2\times\mathbb{Z}_2$
symmetry~\cite{Bhattacharya:2013hva}, $\mathbb{Z}_2\times U(1)_{PQ}$
symmetry~\cite{Alves:2016bib}, gauged $U(1)_{B-L}$
symmetry~\cite{Klasen:2016qux} and Kaluza-Klein theory~\cite{Chialva:2012rq}. 

In this paper, we consider a hidden $U(1)_D$ extension of the Standard
Model (SM), which is spontaneously broken to the residual $\mathbb{Z}_4$
symmetry. Due to the remnant symmetry and an assumption of mass
hierarchy, two particles in the dark sector can be stabilized. 
We discuss the two dark matter components in this model. 
The relic density of the lighter component can be
determined by forbidden channels while that for the heavier
component is fixed by thermal freeze-out. As a consequence, 
self-interacting cross section for the lighter dark matter can
be large enough to solve the small scale problems. 
The two dark matter components are closely correlated with each other
and those implications are also discussed. 

This paper is organized as follows. 
In Section~\ref{sec:2}, details of the model are presented, and 
some basic constraints relevant to new particles are discussed. 
The relic density and self-interaction of two-component dark matter are
discussed and numerically evaluated in Section~\ref{sec:3}. 
Section~\ref{sec:4} is devoted to give detection properties of the two
dark matter components. 
Summary and conclusions are given in Section~\ref{sec:5}.

\section{The Model}
\label{sec:2}
We consider the model extended with the hidden $U(1)_D$ gauge
symmetry. The new particle contents and the charge assignments of the
$U(1)_D$ symmetry are shown in Tab.~\ref{tab:particle}. 
The new complex scalar $\Sigma$ which develops a vacuum expectation value (VEV), and
two new inert complex scalars $S$ and $\chi$ are introduced to the SM, 
where the $U(1)_D$ charge of $\Sigma$ is normalized to $1$. 
This normalization would be relevant to perturbativity of the $U(1)_D$
gauge coupling. Namely, the combination of the $U(1)_D$ charge and gauge
coupling is bounded from above,
and fixing the largest $U(1)_D$ charge in the hidden particles to be
one would be reasonable to consider the perturbative gauge coupling. 
This model is automatically anomaly free since we add only new complex
scalars. 
The kinetic terms of the new particles are given by 
\begin{eqnarray}
\mathcal{L}=
\left|D_\mu \Sigma\right|^2
+\left|D_\mu S\right|^2
+\left|D_\mu \chi\right|^2
-\frac{\epsilon}{2}B_{\mu\nu}Z^{\prime\mu\nu},
\label{L_kin}
\end{eqnarray}
where the covariant derivative is defined by
$D_\mu\equiv\partial_\mu+iQ_Dg_DZ'_\mu$ with the $U(1)_D$ charge $Q_D$
given in Tab.~\ref{tab:particle} 
and the $U(1)_D$ gauge coupling constant $g_D$. 
The last term in Eq.(\ref{L_kin}) is the kinetic mixing between the $U(1)_Y$ and $U(1)_D$ gauge
fields which gives the interaction between the SM and
the hidden sector particles. 
In fact, non-zero kinetic mixing is required to ensure that the SM
and the dark sector particles are in thermal (kinetic) equilibrium at
the early universe.\footnote{Thermal equilibrium between the SM and dark
sectors may also be achieved with the couplings in the scalar potential. 
In this case, the extra Higgs boson denoted by $H$ should be
light enough in order to induce a sufficient reaction rate between the SM
and dark sectors. 
} 
The off-diagonal kinetic term given by the kinetic mixing $\epsilon$ can
be diagonalized to obtain the physical mass eigenstates of the gauge bosons. 
The detailed discussion of the diagonalization has
been given in Refs.~\cite{Choi:2015bya, Hochberg:2015vrg} for example. 
In particular, when a light $Z^\prime$ gauge boson ($m_{Z^\prime}\ll
m_Z$) is considered as we will see below, the matrix of the kinetic
terms is diagonalized within good approximation with the following replacement
\begin{eqnarray}
Z_\mu
\hspace{-0.2cm}&\to&\hspace{-0.2cm}
Z_\mu,\\
A_\mu
\hspace{-0.2cm}&\to&\hspace{-0.2cm}
A_\mu-\epsilon_\gamma Z_\mu^\prime,\\
Z_\mu^\prime
\hspace{-0.2cm}&\to&\hspace{-0.2cm}
Z_\mu^\prime-\epsilon_\gamma\tan\theta_W Z_\mu,
\end{eqnarray}
where $\epsilon_\gamma\equiv \epsilon\cos\theta_W$ and $\theta_W$ is the
Weinberg angle. The kinetic mixing $\epsilon_\gamma$ is experimentally
constrained as $\epsilon_\gamma\lesssim10^{-3}$
for $m_{Z^\prime} \sim $ MeV-GeV as we will see later. 

\begin{table}[t]
\centering
\caption{New particle contents and $U(1)_D$ charges $Q_D$. }
\begin{tabular}{ccccc}\bhline{1pt}
                && $\Sigma$    & $S$ & $\chi$\\\hline
$Q_D$  && $1$       & $-1/2$ & $1/4$\\
Remnant $\mathbb{Z}_4$  && $0$       & $2$ & $1$\\
Spin  && $0$       & $0$ & $0$\\
\bhline{1pt}
\end{tabular}
\label{tab:particle}
\end{table}

The full renormalizable scalar potential is written down as
\begin{eqnarray}
\mathcal{V}\hspace{-0.2cm}&=&\hspace{-0.2cm}
\mu_\Phi^2|\Phi|^2+\mu_\Sigma^2|\Sigma|^2+\mu_S^2|S|^2+\mu_\chi^2|\chi|^2
+\frac{\lambda_\Phi}{4}|\Phi|^4+\frac{\lambda_\Sigma}{4}|\Sigma|^4
+\frac{\lambda_S}{4}|S|^4+\frac{\lambda_\chi}{4}|\chi|^4
\nonumber\\
&&\hspace{-0.2cm}
+\lambda_{\Phi\Sigma}|\Phi|^2|\Sigma|^2
+\lambda_{\Phi S}|\Phi|^2|S|^2
+\lambda_{\Phi\chi}|\Phi|^2|\chi|^2
+\lambda_{\Sigma S}|\Sigma|^2|S|^2
+\lambda_{\Sigma\chi}|\Sigma|^2|\chi|^2
+\lambda_{S\chi}|S|^2|\chi|^2
\nonumber\\
&&\hspace{-0.2cm}
+\left(\frac{\kappa}{2}\Sigma
  S^2+\frac{\mu}{2}S\chi^2+\frac{\lambda}{2}\Sigma S{\chi^\dag}^2
+\mathrm{H.c.}\right).
\end{eqnarray}
We assume that only the SM Higgs doublet $\Phi$ and the new complex
scalar $\Sigma$ have VEVs ($\langle\Sigma\rangle\ll\langle\Phi\rangle$)
as denoted by
\begin{equation}
\Phi=\left(
\begin{array}{c}
0\\
\langle\Phi\rangle+\phi^0/\sqrt{2}
\end{array}
\right),\qquad
\Sigma=
\langle\Sigma\rangle+\frac{\sigma}{\sqrt{2}}. 
\end{equation}
The $U(1)_D$ symmetry is spontaneously broken by the VEV of $\Sigma$,
and the mass of $Z'$ gauge boson is generated 
due to the symmetry breaking as
\begin{equation}
m_{Z^\prime}^2=g_D^2\langle\Sigma\rangle^2.
\label{eq:zprime}
\end{equation}
For hierarchical VEVs
$\langle\Sigma\rangle\ll\langle\Phi\rangle$,
a light $Z^\prime$ boson ($m_{Z^\prime}\ll m_Z$) is obtained.
In the scalar potential, the cubic term $(\kappa/2)\Sigma S^2$ induces a mass
splitting between the CP-even and odd states of $S$ after the $U(1)_D$
symmetry breaking. Their masses are given by 
\begin{eqnarray}
m_{s_R}^2
\hspace{-0.2cm}&=&\hspace{-0.2cm}
\mu_S^2+\lambda_{\Phi
 S}\langle\Phi\rangle^2
+\lambda_{\Sigma S}\langle\Sigma\rangle^2
+\kappa\langle\Sigma\rangle,\\
m_{s_I}^2
\hspace{-0.2cm}&=&\hspace{-0.2cm}
\mu_S^2+\lambda_{\Phi S}\langle\Phi\rangle^2
+\lambda_{\Sigma S}\langle\Sigma\rangle^2
-\kappa\langle\Sigma\rangle,
\end{eqnarray}
where $S$ is decomposed as $S=(s_R+is_I)/\sqrt{2}$. 
Hereafter we assume that the parameter $\kappa$ is small, namely 
the mass splitting between $s_R$ and $s_I$ is small enough.
However one should note that the parameter $\kappa$ is
bounded from below. 
It is relevant to the inelastic scattering for direct detection of dark matter
$s_I$.\footnote{The CP-odd scalar $s_I$ is identified as one of the two dark matter
components with the mass below GeV scale.} 
If $\kappa$ is small enough, 
the inelastic scattering with electron via the $Z^\prime$ gauge boson exchange $s_I e^-\to
s_Re^-$ may give a constraint on this model. 
In order to evade this inelastic scattering, the parameter $\kappa$ is constrained as
$\kappa\gtrsim m_{s_I}v^2/2$ 
where $m_{s_I}\approx \langle\Sigma\rangle\gg m_{s_R}-m_{s_I}$ is
assumed. 
Here $v\sim10^{-3}$ is the dark matter velocity in the present universe.
Thus for example, if $m_{s_I}\sim200~\mathrm{MeV}$, the lower bound for $\kappa$ is
given by $\kappa\gtrsim100~\mathrm{eV}$. 
Another lower bound of the parameter $\kappa$ is induced from Big
Bang Nucleosynthesis (BBN). 
This is because the successful BBN is spoiled if the CP-even state $s_R$
has a too long lifetime
($\tau_{s_R}\gtrsim0.1~\mathrm{s}$)~\cite{Kawasaki:2004qu,
Jedamzik:2006xz}. 
Assuming $m_{s_R}-m_{s_I}\gg m_e$, the decay width of $s_R$ can
roughly be computed as 
\begin{equation}
\Gamma_{s_R\to s_Ie^+e^-}\approx
\frac{\epsilon_\gamma^2\alpha_\mathrm{em}}{192\pi^2g_D}\frac{\kappa^3}
{m_{s_I}m_{Z^\prime}}.
\end{equation}
From this formula, fixing $m_{s_I}\approx m_{Z^\prime}/2$,
$g_D=1$, $\epsilon_\gamma=10^{-7}$ for example, 
the lower bound of $\kappa$ is derived as
\begin{equation}
\kappa\gtrsim 10~\mathrm{MeV}
\times\left(\frac{m_{Z^\prime}}{100~\mathrm{MeV}}\right)^{2/3}.
\end{equation}
Therefore one can see that the BBN constraint is stronger and we
take the parameter $\kappa$ satisfying the BBN constraint in the
following discussion. 
When the relic density of the dark matter is computed, we can neglect the mass
difference between $s_R$ and $s_I$, and thus $S$ is regarded as a
complex scalar particle because $s_R$ and $s_I$ are thermalized in the
early universe. 
On the other hand, the mass difference cannot be neglected when detection
properties such as direct and indirect detection rates are computed in the current universe. 

The mass of $\chi$ is given by
\begin{equation}
m_\chi^2=\mu_\chi^2+\lambda_{\Phi\chi}\langle\Phi\rangle^2+
\lambda_{\Sigma\chi}\langle\Sigma\rangle^2.
\end{equation}
The neutral component of the SM Higgs doublet $\phi^0$ and the scalar
$\sigma$ in the dark sector mix with each other, and the mass matrix can be diagonalized as
\begin{eqnarray}
\mathcal{V}
\hspace{-0.2cm}&\supset&\hspace{-0.2cm}
\frac{1}{2}\left(
\begin{array}{cc}
\phi^0 & \sigma
\end{array}
\right)\left(
\begin{array}{cc}
\lambda_\Phi\langle\Phi\rangle^2 & 
2\lambda_{\Phi\Sigma}\langle\Phi\rangle\langle\Sigma\rangle\\
2\lambda_{\Phi\Sigma}\langle\Phi\rangle\langle\Sigma\rangle & 
\lambda_\Sigma\langle\Sigma\rangle^2
\end{array}
\right)\left(
\begin{array}{c}
\phi^0\\
\sigma
\end{array}
\right)\nonumber\\
\hspace{-0.2cm}&=&\hspace{-0.2cm}
\frac{1}{2}\left(
\begin{array}{cc}
h & H
\end{array}
\right)\left(
\begin{array}{cc}
m_h^2 & 0\\
0 & m_H^2
\end{array}
\right)\left(
\begin{array}{c}
h\\ H
\end{array}
\right),
\label{eq:higgs}
\end{eqnarray}
where the minimum conditions of the scalar potential are imposed. 
The mass eigenstates $h$
and $H$ are understood as the SM-like Higgs 
boson with $m_h=125~\mathrm{GeV}$ and an extra (dark) Higgs boson, respectively.
The gauge eigenstates $\phi^0$ and $\sigma$ can be rewritten by the mass eigenstates as
\begin{equation}
\left(
\begin{array}{c}
\phi^0\\
\sigma
\end{array}
\right)=\left(
\begin{array}{cc}
\cos\alpha & -\sin\alpha\\
\sin\alpha & \cos\alpha
\end{array}
\right)\left(
\begin{array}{c}
h\\ H
\end{array}
\right)\quad\text{with}\quad
\sin2\alpha=\frac{4\lambda_{\Phi\Sigma}\langle\Phi\rangle\langle\Sigma\rangle}{m_h^2-m_H^2}.
\end{equation}
The mixing angle $\sin\alpha$ is constrained by the electroweak
precision data, Higgs coupling measurements and direct search for a
new scalar, and 
the current bound is given by $\sin\alpha\lesssim0.01$ for
$m_H\lesssim5~\mathrm{GeV}$~\cite{Robens:2015gla, Falkowski:2015iwa}. 
In this mass region, the strongest bound is given by the decay mode
$B\to K\ell\ell$~\cite{Wei:2009zv, Aaij:2012vr, Lees:2012iw, Schmidt-Hoberg:2013hba}. 
Since the mass of the extra Higgs boson $H$ is basically given by
$m_H^2\sim 4\lambda_{\Phi\Sigma}^2\langle\Sigma\rangle^2/\lambda_{\Phi}$
where
$\lambda_\Sigma\langle\Sigma\rangle\ll\lambda_{\Phi\Sigma}\langle\Phi\rangle$
is assumed, 
the scale of $m_H$ is correlated with $m_{Z^\prime}$ like
$m_H^2/m_{Z^\prime}^2\sim 4\lambda_{\Phi\Sigma}^2/(\lambda_\Phi g_D^2)$. 
Therefore, if the gauge coupling is taken as
$g_D\sim 1$, the mass $m_H$ cannot be much larger than $m_{Z^\prime}$
due to the perturbativity of the couplings.


The invisible decay of the SM-like Higgs boson $h$ also gives a constraint on the 
quartic couplings in the scalar potential.
The current upper bound of the branching fraction into the invisible decay
mode is given as
$\mathrm{Br}(h\to\mathrm{inv})\leq0.28$~\cite{Aad:2015txa} by the ATLAS 
Collaboration and
$\mathrm{Br}(h\to\mathrm{inv})\leq0.24$~\cite{Khachatryan:2016whc} by
the CMS Collaboration at $95\%$ confidence level. 
If the value of the ATLAS Collaboration is taken as a conservative
limit, this upper bound can be translated into the upper bound of the
invisible decay width as $\Gamma_\mathrm{inv}\leq 1.6~\mathrm{MeV}$
where 
$\Gamma_h^\mathrm{SM}=4.1~\mathrm{MeV}$ is used~\cite{totaldecaywidth}.
In this model, the possible invisible decay channels are given by $h\to
HH,s_Rs_R,s_Is_I, (\chi\chi^\dag)$ and each decay width is computed as
\begin{eqnarray}
\Gamma_{HH}
\hspace{-0.2cm}&=&\hspace{-0.2cm}
\frac{\mu_{hHH}^2}{32\pi
 m_h}\sqrt{1-\frac{4m_H^2}{m_h^2}},\\
\Gamma_{SS^\dag}
\hspace{-0.2cm}&=&\hspace{-0.2cm}
\frac{\mu_{hSS^\dag}^2}{16\pi
 m_h}\sqrt{1-\frac{4m_S^2}{m_h^2}},\\
\Gamma_{\chi\chi^\dag}
\hspace{-0.2cm}&=&\hspace{-0.2cm}
\frac{\mu_{h\chi\chi^\dag}^2}{16\pi
 m_h}\sqrt{1-\frac{4m_\chi^2}{m_h^2}},
\end{eqnarray}
with
\begin{eqnarray}
\mu_{hHH}
\hspace{-0.2cm}&\equiv&\hspace{-0.2cm}
\frac{\sin\alpha\cos\alpha}{\sqrt{2}}\Bigl(
\lambda_{\Phi}\langle\Phi\rangle\sin\alpha+\lambda_{\Sigma}\langle\Sigma\rangle\cos\alpha
\Bigr)\nonumber\\
\hspace{-0.2cm}&&\hspace{-0.2cm}
+\sqrt{2}\lambda_{\Phi\Sigma}\Bigl(
\langle\Phi\rangle\cos\alpha(1-3\sin^2\alpha)
+\langle\Sigma\rangle\sin\alpha(1-3\cos^2\alpha)
\Bigr),\\
\mu_{hSS^\dag}
\hspace{-0.2cm}&\equiv&\hspace{-0.2cm}
\sqrt{2}\Bigl(\lambda_{\Phi S}\langle\Phi\rangle\cos\alpha
+\lambda_{\Sigma S}\langle\Sigma\rangle\sin\alpha\Bigr),
\label{eq:SSh}\\
\mu_{h\chi\chi^\dag}
\hspace{-0.2cm}&\equiv&\hspace{-0.2cm}
\sqrt{2}\Bigl(\lambda_{\Phi\chi}\langle\Phi\rangle\cos\alpha
+\lambda_{\Sigma\chi}\langle\Sigma\rangle\sin\alpha \Bigr), 
\end{eqnarray}
where since the CP-even and odd states $s_R$ and $s_I$ are nearly
degenerate, these contributions are approximated by $\Gamma_{SS^\dag}$. 
Thus if all the hidden particles are much lighter than the SM-like Higgs
boson, the couplings are constrained as
\begin{equation}
\mu_{hHH}^2+2\mu_{hSS^\dag}^2+2\mu_{h\chi\chi^\dag}^2\lesssim 4.5~\mathrm{GeV}. 
\end{equation}
In the following analysis, we discuss parameter regions in which the above constraints are satisfied.

In this model, a remnant $\mathbb{Z}_4$ symmetry remains after the
$U(1)_D$ symmetry breaking as one can see from Tab.~\ref{tab:particle}. 
Due to this $\mathbb{Z}_4$ symmetry, the decay of $\chi$ is
forbidden and thus $\chi$ can be a dark matter candidate. 
Moreover, depending on the mass hierarchy, we may have
a second dark matter component $s_I$ because the decay
of $s_I$ is forbidden if $m_{s_I}\leq 2m_\chi$ is satisfied.
In the following, we consider two-component dark matter composed of $s_I$ and $\chi$ with
the mass hierarchy $
m_{s_I}(\approx m_{s_R})
\lesssim m_{Z^\prime}, m_H< m_\chi$. 
In particular, we are interested in the mass region of
$m_{s_I}\lesssim\mathcal{O}(100)~\mathrm{MeV}$ and
$m_\chi\gtrsim1~\mathrm{GeV}$ for large self-interaction
induced by the lighter component $s_I$.
There are two kinds of interactions between dark matter and the SM
particles which are the gauge coupling $g_D$ and the Higgs couplings in
the scalar potential. 
Both interactions would play an important role for dark matter phenomenology.
Note that a certain degree of parameter tuning is needed to achieve the
hierarchical masses between the two dark matter components in this
model.\footnote{Although it is difficult to derive theoretically such a
large mass hierarchy within this model, 
it may be achieved in different frameworks such as
global $U(1)$ models in which the pseudo-Goldstone boson can be identified 
as a lighter dark matter candidate corresponding to $s_I$. 
} 
The magnitude of the tuning would be the same order as that in the SM in
which one 
needs a tuning in order to obtain the hierarchical fermion masses.


\section{Dark Matter}
\label{sec:3}
\subsection{Relic Density}
In general, the coupled Boltzmann equations for two-component dark matter
$S$ and $\chi$ should be solved in order to compute the relic densities
since in addition to the standard annihilation processes, various processes
such as co-annihilations, semi-annihilations and conversion processes
between two dark matter components should be taken into
account.\footnote{As mentioned above, the complex scalar $S$ can be
effectively regarded as a dark matter particle in the early universe since the mass
splitting between $s_R$ and $s_I$ is very small.} 
For example, the semi-annihilation processes $\chi\chi\to SH,Sh$ occur if
the quartic coupling $\lambda$ is large enough, however
these processes and the other semi-annihilation processes can be
neglected by assuming small couplings $\kappa$, $\mu$ and $\lambda$. 
Since we focus on a large mass hierarchy between two dark matter
species ($m_S\ll m_\chi$), 
the equations are almost decoupled and given by
\begin{eqnarray}
\frac{dn_S}{dt}+3Hn_S
\hspace{-0.2cm}&=&\hspace{-0.2cm}
-\frac{1}{2}\langle\sigma_S{v}\rangle
\Bigl[
n_S^2-{n_S^\mathrm{eq}}^2
\Bigr],
\label{eq:boltz1}\\
\frac{dn_\chi}{dt}+3Hn_\chi
\hspace{-0.2cm}&=&\hspace{-0.2cm}
-\frac{1}{2}
\langle\sigma_\chi{v}\rangle
\Bigl[
n_\chi^2-{n_\chi^\mathrm{eq}}^2
\Bigr].
\label{eq:boltz2}
\end{eqnarray}
Thus one can independently solve the Boltzmann equations for each dark
matter particle.
The heavier state $\chi$ would firstly decouple from the thermal bath
before the decoupling of the lighter state $S$.
This is naively expected since the mass hierarchy among the two dark matter
particles are sufficiently large. 
The number densities $n_S$ and $n_\chi$ are defined by the total
number densities of ($S$,$S^\dag$) and ($\chi$,$\chi^\dag$), respectively.
Because of this, the factor $1/2$ in front of the cross sections
appears in Eq.~(\ref{eq:boltz1}) and (\ref{eq:boltz2}). 
The thermally averaged cross sections $\langle\sigma_S{v}\rangle$ and
$\langle\sigma_\chi{v}\rangle$ should include all the possible annihilation channels. 
The two dark matter components should have comparable
relic density, 
otherwise this model can be effectively regarded as one-component dark
matter model.
The criterion for the range of the dark matter
fraction is controversial. In this paper, we consider the fraction of each dark matter
component should be larger than $10\%$.

\subsubsection{Relic Density of $S$}
\begin{figure}[t]
\begin{center}
\includegraphics[scale=0.65]{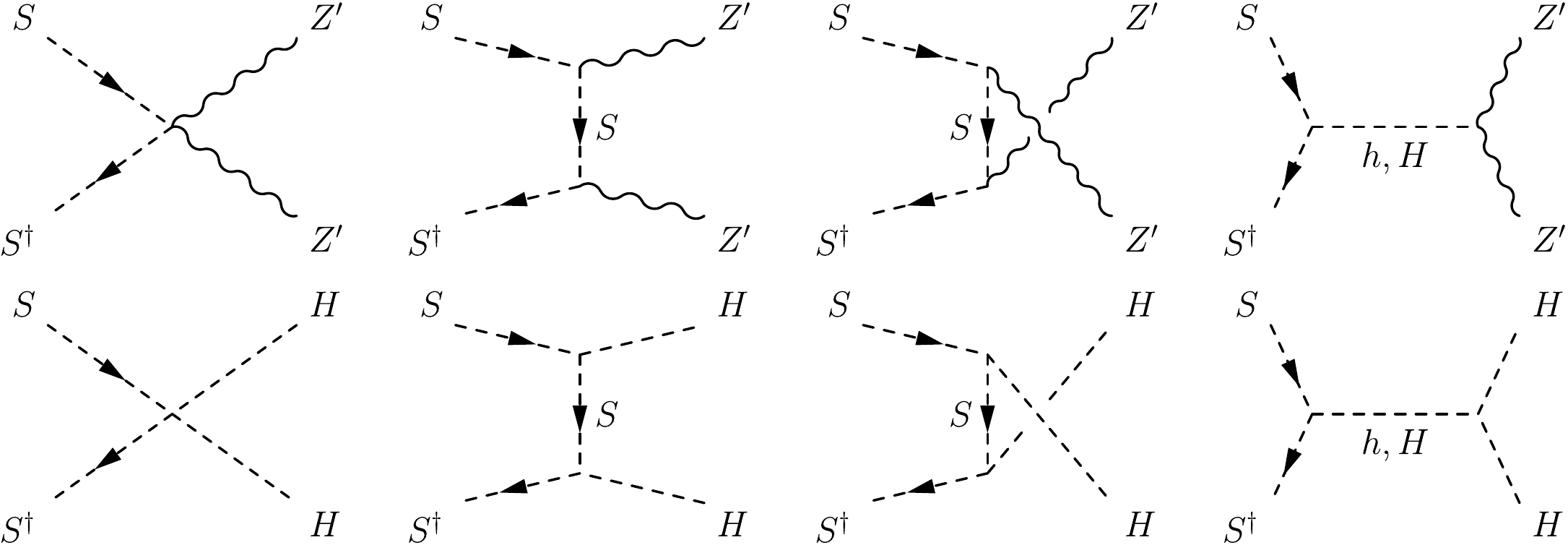}
\caption{Diagrams for the $s_R$, $s_I$ forbidden channels
 ($m_{s_R},m_{s_I}<m_{Z^\prime},m_H$).}
\label{fig:diag1}
\end{center}
\end{figure}

The possible annihilation channel for the lighter dark matter state $S$ is $SS^\dag\to
f\overline{f}$ where $f$ is a SM fermion. However this process is
suppressed by the small kinetic mixing $\epsilon_\gamma\lesssim10^{-3}$
or the Higgs mixing $\sin\alpha$, thus
cannot reproduce the observed relic density.
In addition, since we consider the case of $m_S<m_{Z^\prime},m_H$,
one may think the annihilation processes $SS^\dag\to Z^\prime Z^\prime,HH$ are
kinematically forbidden. However the thermally averaged annihilation
cross sections for these processes are not exactly zero because particles in
the early universe have an energy
distribution like the Maxwell-Boltzmann distribution. 
Thus it is possible to reproduce the correct relic density by the
forbidden channels~\cite{Choi:2016tkj, Griest:1990kh, D'Agnolo:2015koa, Delgado:2016umt}. 
The thermally averaged annihilation cross sections for the forbidden
channels are suppressed by the Boltzmann factor, thus the order of the
magnitude of the couplings required to reproduce the correct relic density would be
larger than the case of normal WIMPs. 

In this model, the relevant diagrams for
the forbidden channels $SS^\dag\to Z^\prime Z^\prime,HH$ are shown in
Fig.~\ref{fig:diag1}. 
Assuming CP invariance, the thermally averaged
annihilation cross section for the process $SS^\dag\to XX$ can be
written with the opposite process 
as~\cite{D'Agnolo:2015koa} 
\begin{equation}
\langle\sigma{v}\rangle_{SS^\dag\to XX}=
\langle\sigma{v}\rangle_{XX\to SS^\dag}
\left(\frac{2n_{X}^\mathrm{eq}}{n_S^\mathrm{eq}}\right)^2=
\langle\sigma{v}\rangle_{XX\to SS^\dag}
\frac{g_X^2m_{X}^3}{m_S^3}e^{-2\Delta_X{z_S}},
\label{eq:forbidden}
\end{equation}
where $X=Z^\prime$ or $H$, $g_X$ is the degrees of freedom for $X$
($g_{Z^\prime}=3$ and $g_H=1$),
$\Delta_X=(m_{X}-m_S)/m_S$ and $z_S=m_S/T$.
The factor $2$ in the middle of Eq.~(\ref{eq:forbidden}) appears because we
defined $n_S$ as the total number density of $S$ and $S^\dag$. 
Each cross section
$\langle\sigma{v}\rangle_{XX\to SS^\dag}$ for $X=Z^\prime$ and $H$ is
given by
\begin{eqnarray}
\langle\sigma{v}\rangle_{Z^\prime Z^\prime\to SS^\dag}
\hspace{-0.2cm}&\approx&\hspace{-0.2cm}
\frac{1}{72\pi m_{Z^\prime}^2}\left(\frac{g_D}{2}\right)^4
\left[
11-24\frac{m_S^2}{m_{Z^\prime}^2}+16\frac{m_S^4}{m_{Z^\prime}^4}
+3\left(C_R^2+C_I^2\right)
\right.\nonumber\\
\hspace{-0.2cm}&&\hspace{-0.2cm}
\hspace{3cm}
\left.
-2C_R\left(1-4\frac{m_S^2}{m_{Z^\prime}^2}\right)
\right]\sqrt{1-\frac{m_S^2}{m_{Z^\prime}^2}},
\label{eq:SStoZpZp}\\
\langle\sigma{v}\rangle_{HH\to SS^\dag}
\hspace{-0.2cm}&\approx&\hspace{-0.2cm}
\frac{1}{32\pi m_H^2}
\left|\lambda_{HHSS^\dag}-\frac{2\mu_{HSS^\dag}^2}{m_H^2}
+\frac{\mu_{HHH}\:\mu_{HSS^\dag}}{3m_H^2+im_H\Gamma_H}\right|^2
\sqrt{1-\frac{m_S^2}{m_H^2}},
\label{eq:SStoHH}
\end{eqnarray}
in the non-relativistic limit ($v\to0$). 
The new parameters in Eq.~(\ref{eq:SStoZpZp}) and (\ref{eq:SStoHH}) are defined by
\begin{eqnarray}
C_R
\hspace{-0.2cm}&\equiv&\hspace{-0.2cm}
\frac{4\sqrt{2}\cos\alpha}{g_D}
\:\frac{\mu_{HSS^\dag}m_{Z^\prime}(4m_{Z^\prime}^2-m_H^2)}
{(4m_{Z^\prime}^2-m_H^2)^2+m_H^2\Gamma_H^2},\\
C_I
\hspace{-0.2cm}&\equiv&\hspace{-0.2cm}
-\frac{4\sqrt{2}\cos\alpha}{g_D}
\:\frac{\mu_{HSS^\dag}m_{Z^\prime}m_H\Gamma_H}
{(4m_{Z^\prime}^2-m_H^2)^2+m_H^2\Gamma_H^2},\\
\lambda_{HHSS^\dag}
\hspace{-0.2cm}&\equiv&\hspace{-0.2cm}
\lambda_{\Phi S}\sin^2\alpha+\lambda_{\Sigma S}\cos^2\alpha,\\
\mu_{HSS^\dag}
\hspace{-0.2cm}&\equiv&\hspace{-0.2cm}
\sqrt{2}
\Bigl(-\lambda_{\Phi S}\langle\Phi\rangle\sin\alpha
+\lambda_{\Sigma S}\langle\Sigma\rangle\cos\alpha\Bigr),
\label{eq:HSS}
\\
\mu_{HHH}
\hspace{-0.2cm}&\equiv&\hspace{-0.2cm}
-\frac{3}{\sqrt{2}}\Bigl(-\lambda_{\Phi}\langle\Phi\rangle\sin^3\alpha
+\lambda_{\Sigma}\langle\Sigma\rangle\cos^3\alpha\Bigr)\nonumber\\
\hspace{-0.2cm}&&\hspace{-0.2cm}
+\frac{6}{\sqrt{2}}\lambda_{\Phi\Sigma}\sin\alpha\cos\alpha
\Bigl(
-\langle\Phi\rangle\cos\alpha+\langle\Sigma\rangle\sin\alpha
\Bigr).
\end{eqnarray}
The decay width for the extra Higgs boson $H$ is given by
$
\Gamma_H=\Gamma_{Z^\prime
 Z^\prime}+\Gamma_{SS^\dag}+\Gamma_{e\overline{e}}~(+\Gamma_{\mu\overline{\mu}}). 
$
Each decay width is computed as
\begin{eqnarray}
\Gamma_{f\overline{f}}
\hspace{-0.2cm}&=&\hspace{-0.2cm}
\frac{y_f^2\sin^2\alpha m_H}{16\pi}\left(1-4\frac{m_f^2}{m_H^2}\right)^{3/2},\\
\Gamma_{Z^\prime Z^\prime}
\hspace{-0.2cm}&=&\hspace{-0.2cm}
\frac{g_D^2\cos^2\alpha m_{Z^\prime}^2}{4\pi m_H}
\left(3-\frac{m_H^2}{m_{Z^\prime}^2}+\frac{1}{4}\frac{m_H^4}{m_{Z^\prime}^4}\right)
\sqrt{1-4\frac{m_{Z^\prime}^2}{m_H^2}},\\
\Gamma_{SS^\dag}
\hspace{-0.2cm}&\approx&\hspace{-0.2cm}
\frac{\mu_{HSS^\dag}^2}{16\pi m_H}\sqrt{1-4\frac{m_S^2}{m_H^2}},
\end{eqnarray}
where $y_f$ is the SM Yukawa coupling for the fermion $f$. 
The relic density of the lighter dark matter particle $S$ can be
determined by solving the Boltzmann equation in Eq.~(\ref{eq:boltz1})
with sum of these cross sections for the forbidden channels. 

In the above computation of the relic density, the thermal equilibrium
between the dark sector and the SM sector is implicitly assumed. 
The condition for (kinetic) thermal equilibrium is given by
$\Gamma_\mathrm{kin}>H$ at the freeze-out temperature of dark matter
$s_I$ where
$H$ is the Hubble parameter and $\Gamma_\mathrm{kin}$ is defined by 
$\Gamma_\mathrm{kin}\equiv\langle\sigma_\mathrm{el}{v}\rangle
n_\mathrm{SM}$ with the elastic scattering cross section
$\sigma_\mathrm{el}v$ and the number density of the SM particles $n_\mathrm{SM}$. 
The most relevant scattering process would be $s_Ie^\pm\to s_Ie^\pm$ or $s_I\mu^\pm\to s_I\mu^\pm$
through the $t$-channel diagram mediated by the extra Higgs boson $H$
depending on the dark matter mass $m_{s_I}$. 
Since the cross section is suppressed by the small mixing angle
$\sin\alpha$ and the electron Yukawa coupling
for the case of scattering with $e^\pm$, 
the scattering cross section $\sigma_\mathrm{el}v$ could be too small to
keep the thermal equilibrium. 
If the condition for thermal equilibrium is not satisfied, 
the temperature of the dark sector would differentiate from
that of the SM sector, and it could make the above computation change if
the difference of the temperature is large enough. 

In some parameter space, the 3-to-2 
annihilation processes like $SSS^\dag\to 
SZ^\prime$ may be relevant in the mass range we focus on. 
Such a process may compete with the forbidden channels discussed above. 
If the 3-to-2 annihilation processes are dominant and the relic
density is determined by these processes, 
the dark matter particle $S$ would be identified as a so-called SIMP
candidate~\cite{Hochberg:2014dra, Hochberg:2014kqa, Bernal:2015bla,
Choi:2015bya, Bernal:2015xba, Choi:2016hid}. 
Since considering a SIMP dark matter candidate in this model requires a
certain degree of parameter tuning so that the 3-to-2 processes become
dominant, we do not consider this possibility.

\subsubsection{Relic Density of $\chi$}
\begin{figure}[t]
\begin{center}
\includegraphics[scale=0.65]{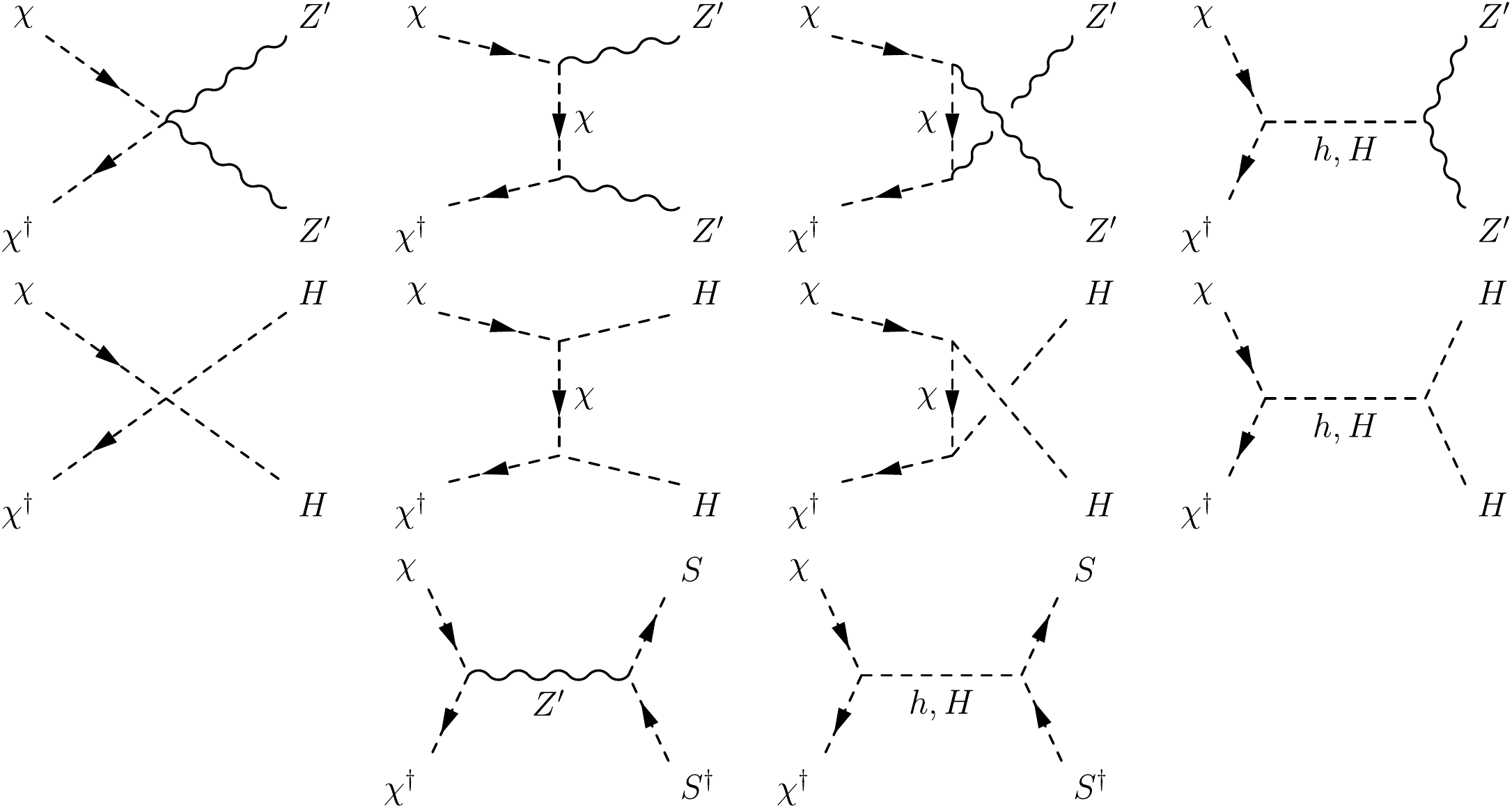}
\caption{Diagrams for the $\chi$ annihilation.}
\label{fig:diag2}
\end{center}
\end{figure}

For the heavier dark matter particle $\chi$, 
three annihilation channels into the particles in the dark sector
$\chi\chi^\dag\to Z^\prime Z^\prime, SS^\dag, HH$ exist. 
The relevant diagrams for the annihilation channel $\chi\chi^\dag\to
Z^\prime Z^\prime$ are shown in the top of Fig.~\ref{fig:diag2}.
The annihilation cross section is given by
\begin{equation}
\sigma{v}_{\chi\chi^\dag\to Z^\prime Z^\prime}
\approx
\frac{1}{16\pi m_\chi^2}\left(\frac{g_D}{4}\right)^4
\frac{8m_\chi^4-8m_\chi^2m_{Z^\prime}^2+3m_{Z^\prime}^4}{(m_{Z^\prime}^2-2m_\chi^2)^2}
\sqrt{1-\frac{m_{Z^\prime}^2}{m_\chi^2}},
\label{eq:sv2}
\end{equation}
in the non-relativistic limit ($v\to0$) where the
contribution of the $s$-channel diagrams mediated by the Higgs bosons
$h,H$ can be neglected with small couplings
$\lambda_{\Phi\chi},\lambda_{\Sigma\chi}\ll1$. 
In fact, such small couplings are required to evade the strong constraint of
the dark matter direct detection experiments. 
The annihilation channel
$\chi\chi^\dag\to HH$ 
in the second line of Fig.~\ref{fig:diag2} 
and  the channels into the SM particles 
also exist.
These may affect to the
computation of the relic density of $\chi$, however 
these channels can be regarded as sub-dominant compared to the channel
$\chi\chi^\dag\to Z^\prime Z^\prime$ due to $\lambda_{\Phi\chi},\lambda_{\Sigma\chi}\ll1$. 
The additional annihilation process $\chi\chi^\dag\to SS^\dag$ shown in
the bottom of Fig.~\ref{fig:diag2} may also be relevant, and 
the annihilation cross section for this process can be computed as
\begin{equation}
\sigma{v}_{\chi\chi^\dag\to SS^\dag}\approx\frac{1}{6\pi
 m_\chi^2}\left(\frac{g_D}{4}\right)^4\left(1-\frac{m_S^2}{m_\chi^2}\right)^{3/2}
\frac{m_\chi^4 v^2}{(4m_\chi^2-m_{Z^\prime}^2)^2+m_{Z^\prime}^2\Gamma_{Z^\prime}^2}.
\end{equation}
However as one can see from the formula, the annihilation cross section is
suppressed by the dark matter relative velocity $v$, thus this
contribution to the total annihilation 
cross section would be sub-dominant. 

For the heavier dark matter $\chi$, one should take into account the
non-perturbative effect which is so-called Sommerfeld effect if the $Z^\prime$
gauge boson is much lighter than the dark matter particle $\chi$ as in
our case. 
In this case, the wave function of the two dark matter particles in the
initial state is distorted by long-range force, and 
the annihilation cross section would be enhanced~\cite{Hisano:2006nn,
Hisano:2002fk, Hisano:2003ec, Hisano:2004ds, Hisano:2005ec}. 
The Sommerfeld enhancement factor can be obtained by solving the
Schr\"{o}dinger equation for the two body dark matter state $\psi(r)$ which
is given by
\begin{equation}
-\frac{1}{m_\chi}\frac{d^2\psi}{dr^2}+V\psi=\frac{m_\chi
 v^2}{4}\psi,\quad
\text{where}
\quad
V=-\frac{\alpha_{Z^\prime}}{16r}e^{-m_{Z^\prime}r},
\end{equation}
with $\alpha_{Z^\prime}=g_D^2/(4\pi)$. 
Here $r$ is the distance between the two dark matter particles. 
This Schr\"{o}dinger equation is solved under the boundary condition
$d\psi/dr(\infty)=0$ and $\psi(0)=1$, and then the Sommerfeld factor is given by
$S_F\equiv|\psi(\infty)|^2$. 
An approximate analytic solution for the Schr\"{o}dinger equation is
given by~\cite{Feng:2010zp}
\begin{equation}
S_F=\frac{\pi}{16\xi_v}\frac{\sinh\left(\frac{2\pi \xi_v}{\pi^2\xi_{Z^\prime}/6}\right)}
{\cosh\left(\frac{2\pi \xi_v}{\pi^2\xi_{Z^\prime}/6}\right)
-\cos\left(2\pi\sqrt{\frac{1}{16\pi^2\xi_{Z^\prime}/6}-
\frac{\xi_v^2}{(\pi^2\xi_{Z^\prime}/6)^2}}\right)},
\end{equation}
where $\xi_v=v/(2\alpha_{Z^\prime})$ and
$\xi_{Z^\prime}=m_{Z^\prime}/(\alpha_{Z^\prime}m_\chi)$.
We use this formula in numerical calculations below.
Thus the thermally average annihilation cross section for the channel $\chi\chi^\dag\to
Z^\prime Z^\prime$ with the Sommerfeld effect is given by
\begin{equation}
\langle\sigma{v}\rangle_{\chi\chi^\dag\to Z^\prime Z^\prime}=
\frac{z_\chi^{3/2}}{2\sqrt{\pi}}\int_{0}^{\infty}
\left(\sigma{v}_{\chi\chi^\dag\to Z^\prime Z^\prime}\right)
S_Fv^2e^{-\frac{z_\chi
v^2}{4}}dv,
\label{eq:thermal}
\end{equation}
where $z_\chi=m_\chi/T$. 
One can see that when $S_F=1$ in Eq.~(\ref{eq:thermal}), the formula without the
Sommerfeld effect is recovered~\cite{Griest:1990kh}.

\subsection{Self-interacting Cross Section}

Strong self-interaction of dark matter is required in order to solve the
small scale problems: 
cusp-core, too-big-to-tail and missing satellites problems. 
The required magnitude of the self-interacting cross section
is quite large as
$\sigma/m\sim0.5-50~\mathrm{cm^2/g}$~\cite{Elbert:2014bma}. 
Reproducing such a large self-interacting cross section may be difficult for
the dark matter with above electroweak scale mass. 
However it is possible to achieve it if the mass of dark matter is below
GeV scale, and we have such a candidate $s_I$ in this model. 

\begin{figure}[t]
\begin{center}
\includegraphics[scale=0.75]{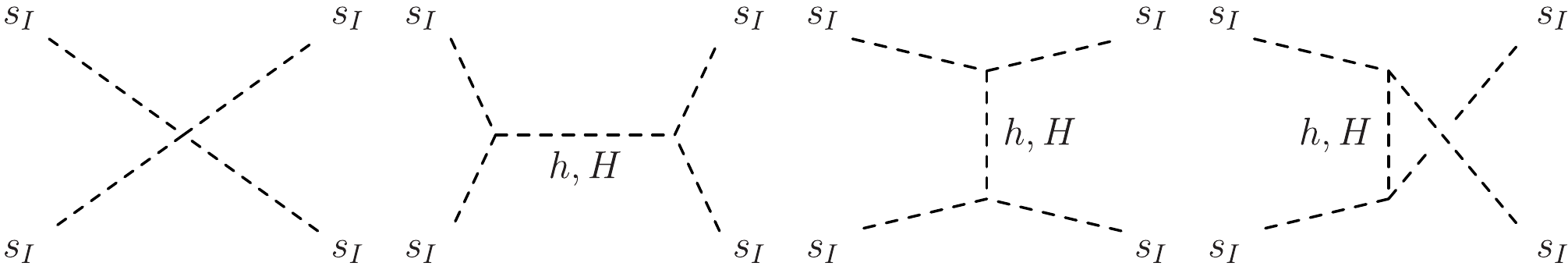}
\caption{Diagrams for the self-interaction process $s_Is_I\to s_Is_I$.}
\label{fig:self}
\end{center}
\end{figure}

The self-interacting cross section for $s_I$ is given by the process
$s_Is_I\to s_Is_I$ whose complete diagrams are shown
in Fig.~\ref{fig:self}, and is computed as
\begin{eqnarray}
\sigma_{s_Is_I\to s_Is_I}
\hspace{-0.2cm}&\approx&\hspace{-0.2cm}
\frac{1}{128\pi m_{s_I}^2}
\left|
\frac{3}{2}\lambda_S
-\frac{2\mu_{Hs_Is_I}^2}{m_H^2}
+\frac{\mu_{Hs_Is_I}^2}{4m_{s_I}^2-m_H^2+im_H\Gamma_H}
\right|^2,
\label{eq:self_res}
\end{eqnarray}
where $\mu_{Hs_Is_I}$ is given by $\mu_{HSS^\dag}$ in Eq.~(\ref{eq:HSS}) for
$\kappa\approx0$, and the contribution of the SM-like Higgs boson is neglected.
Note that only the Higgs couplings contribute to the self-interacting
cross section at the tree level.
The gauge coupling does not give a contribution because it gives only an
inelastic scattering cross section $s_Is_I\to s_Rs_R$.

Since two dark matter components exist in this model, 
the required value of the self-interacting cross section for solving the
small scale problems is scaled by the fraction of the $s_I$ component in the total 
dark matter relic density. 
Thus it would be convenient to define the effective self-interacting
cross section with 
\begin{equation}
\sigma_\mathrm{self}^\mathrm{eff}
=
\left(\frac{\Omega_{s_I}}{\Omega_\mathrm{exp}}\right)^2
\sigma_{s_Is_I\to s_Is_I},
\end{equation}
where the parameters $\Omega_{s_I}$
and $\Omega_\mathrm{exp}\approx0.12/h^2$ ($h$ is the dimensionless Hubble
constant at the current time) are the relic density of $s_I$ and the
experimentally observed total dark matter relic density, respectively. 

The self-interacting cross section for the heavier dark matter state
$\chi$ can also be computed in the same way as the dark matter $s_I$. 
Since the cross section decreases as the dark matter mass increases,
the cross section for the heavier state $\chi$ is expected
to be small. 
Even if the mass of $\chi$ is larger than the electroweak
scale, the self-interacting cross section may be enhanced by the
Sommerfeld effect~\cite{Aarssen:2012fx, Tulin:2012wi}.
However since the effect is not so important for the parameter space we
are interested in, we neglect it.

\subsection{Numerical Computations}

\begin{figure}[t]
\begin{center}
\includegraphics[scale=0.65]{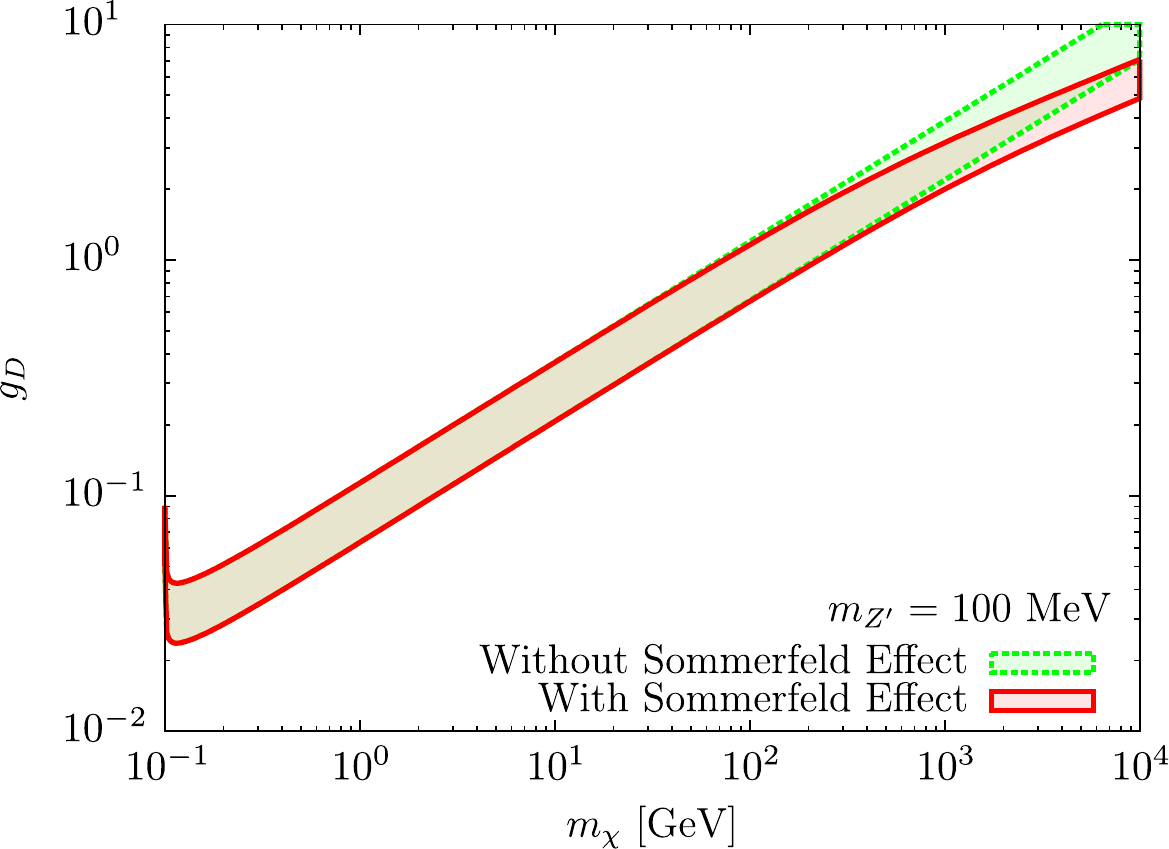}
\caption{Parameter space reproducing the relic density of $\chi$ within
 $0.1<f_\chi<0.9$ in the plane ($m_\chi$, $g_D$) where $m_{Z^\prime}$ is
 fixed to be $m_{Z^\prime}=100~\mathrm{MeV}$ as an example.}
\label{fig:num0}
\end{center}
\end{figure}

\begin{figure}[t]
\begin{center}
\includegraphics[scale=0.65]{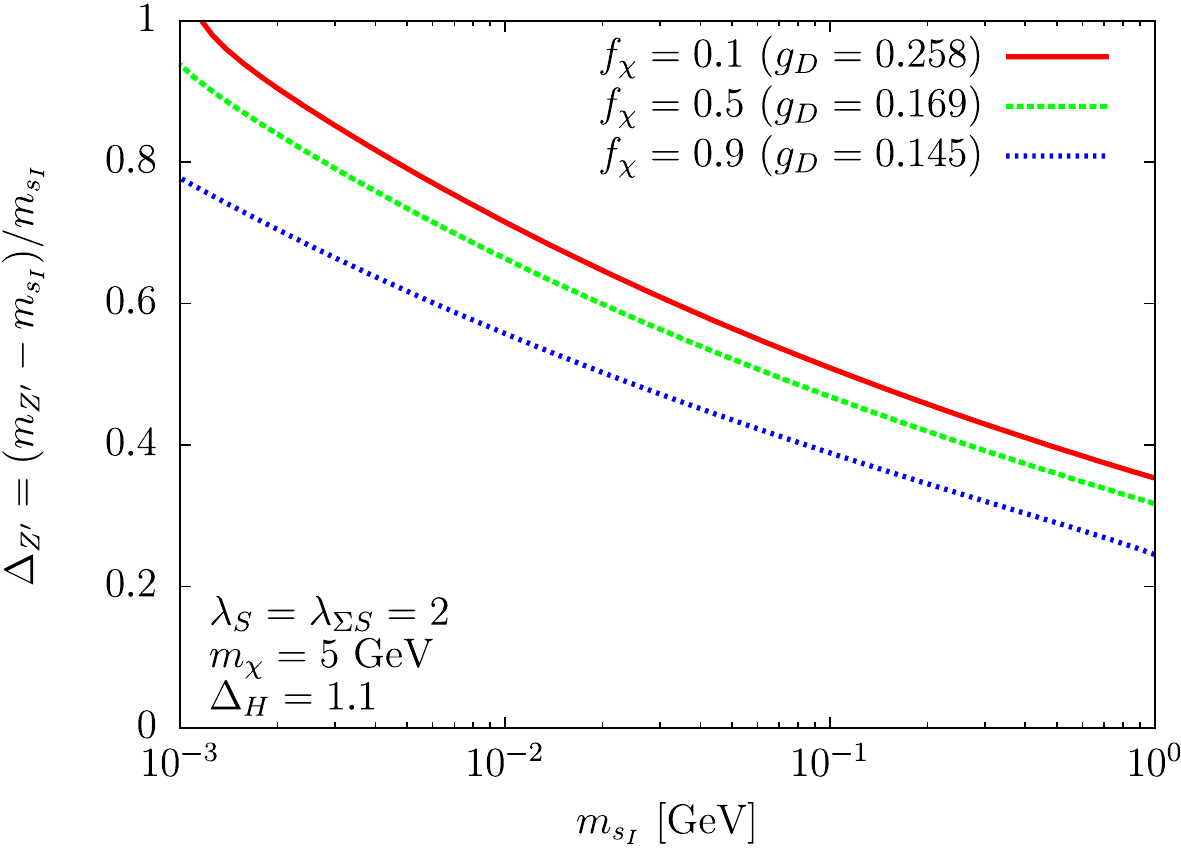}
\includegraphics[scale=0.65]{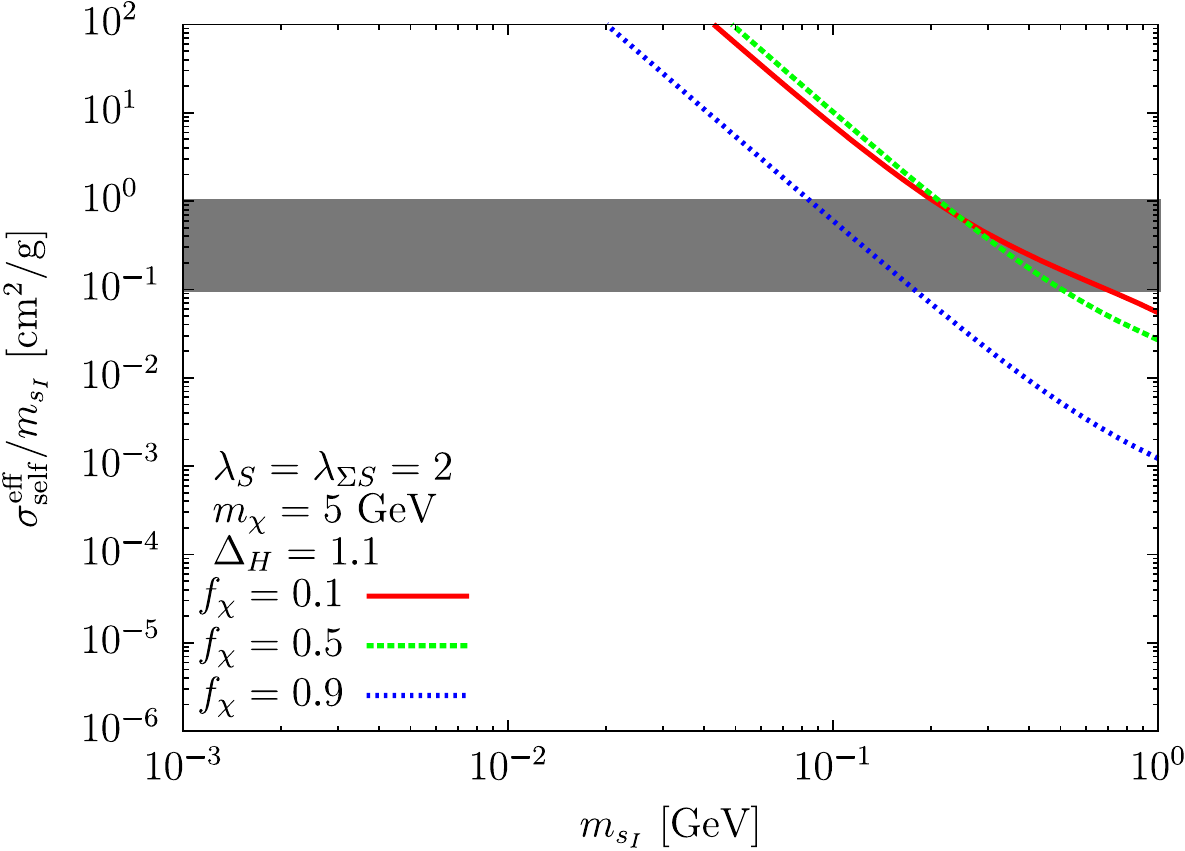}\\
\includegraphics[scale=0.65]{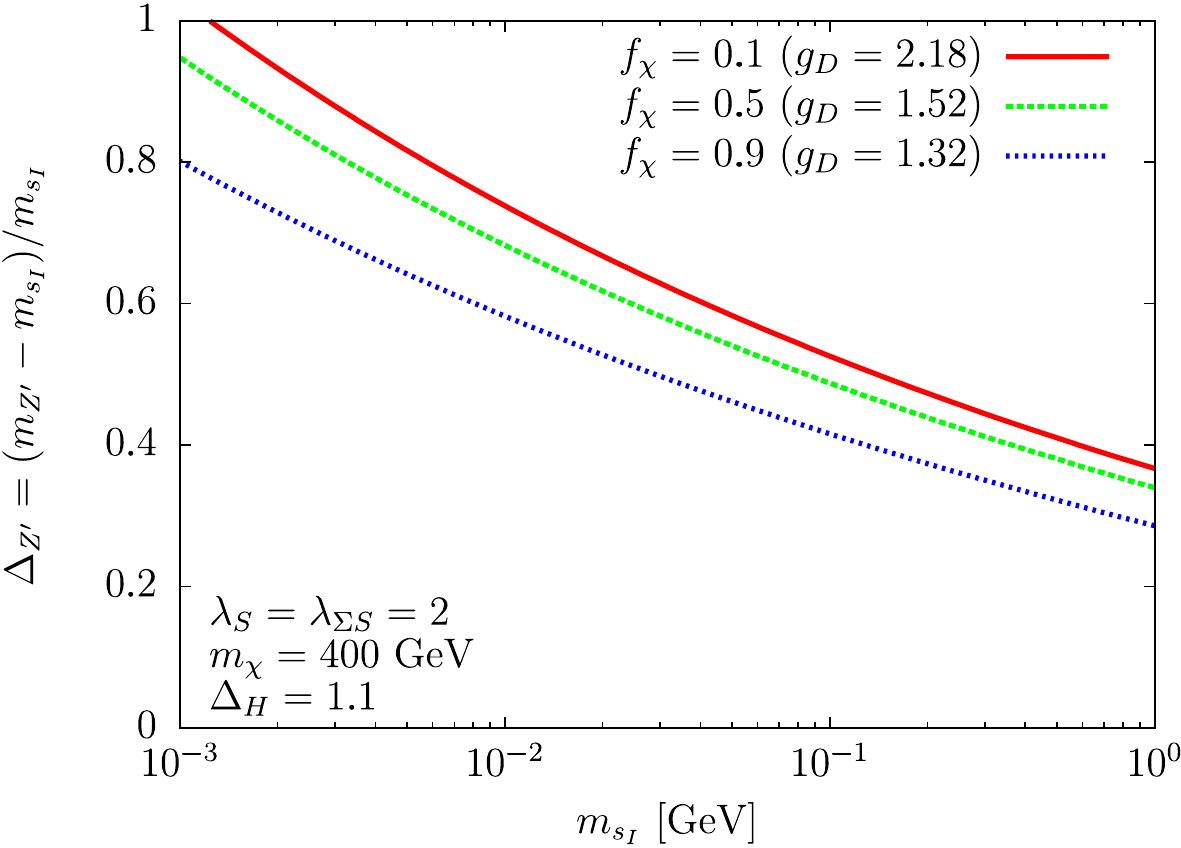}
\includegraphics[scale=0.65]{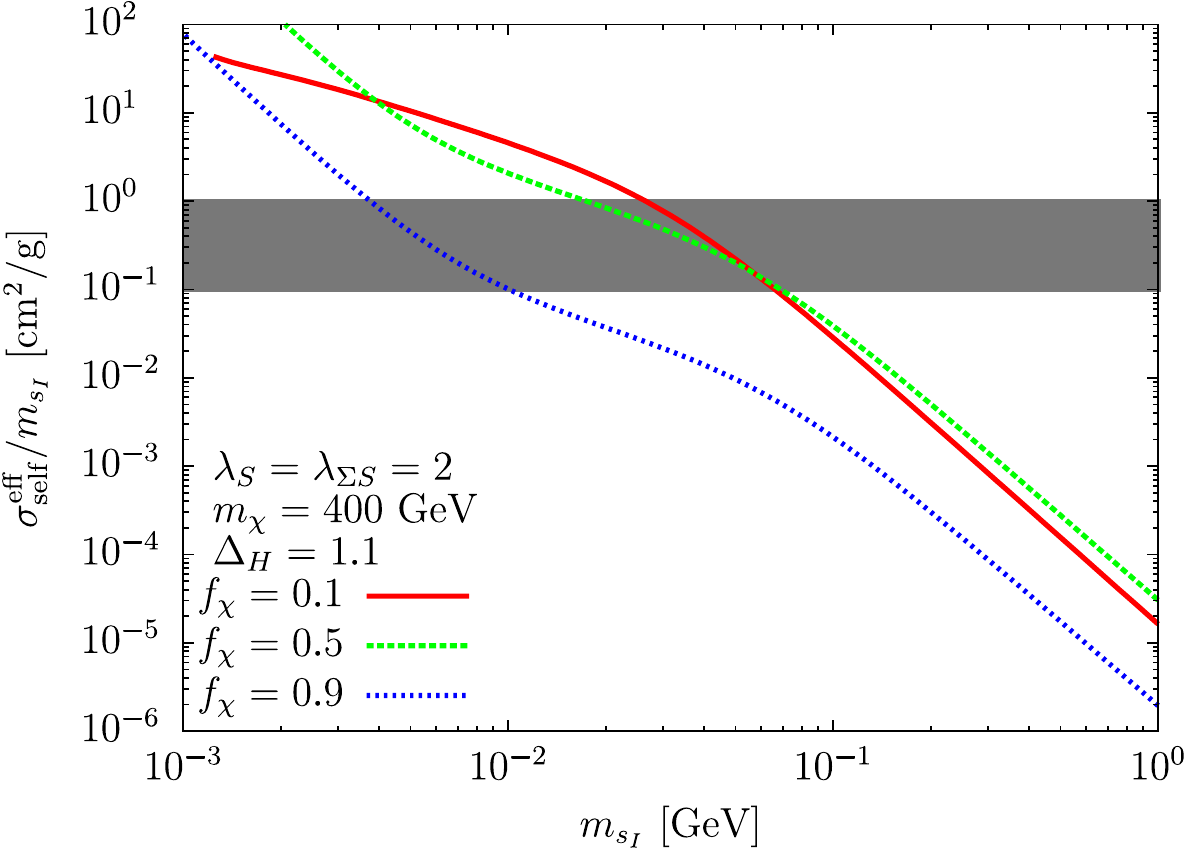}
\caption{(left panels): Contours of the parameters satisfying the dark
 matter relic density in the ($m_{s_I}$, $\Delta_{Z^\prime}$) plane. For each colored
 line, the fraction of two dark matter components are fixed as shown in
 the plots. 
 The relevant parameters are fixed to be $\lambda_{\Sigma
 S}=\lambda_{S}=2$, $\Delta_H=1.1$ and
 $m_\chi=5~\mathrm{GeV}$ (upper plots) or $400~\mathrm{GeV}$ (lower plots). 
 (right panels): The effective self-interacting cross section as a
 function of $m_{s_I}$ where the colored lines correspond to the same
 colored lines in the left plots.}
\label{fig:num1}
\end{center}
\end{figure}

\begin{figure}[t]
\begin{center}
\includegraphics[scale=0.65]{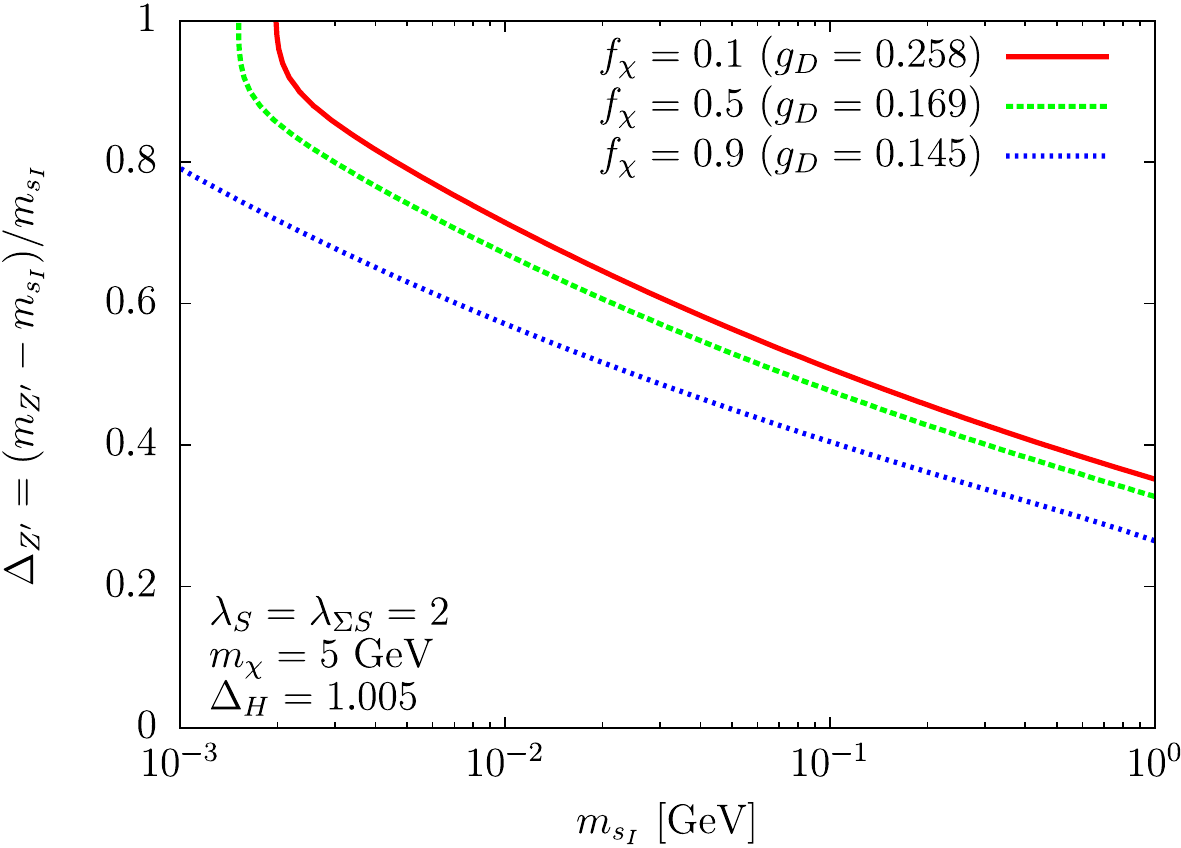}
\includegraphics[scale=0.65]{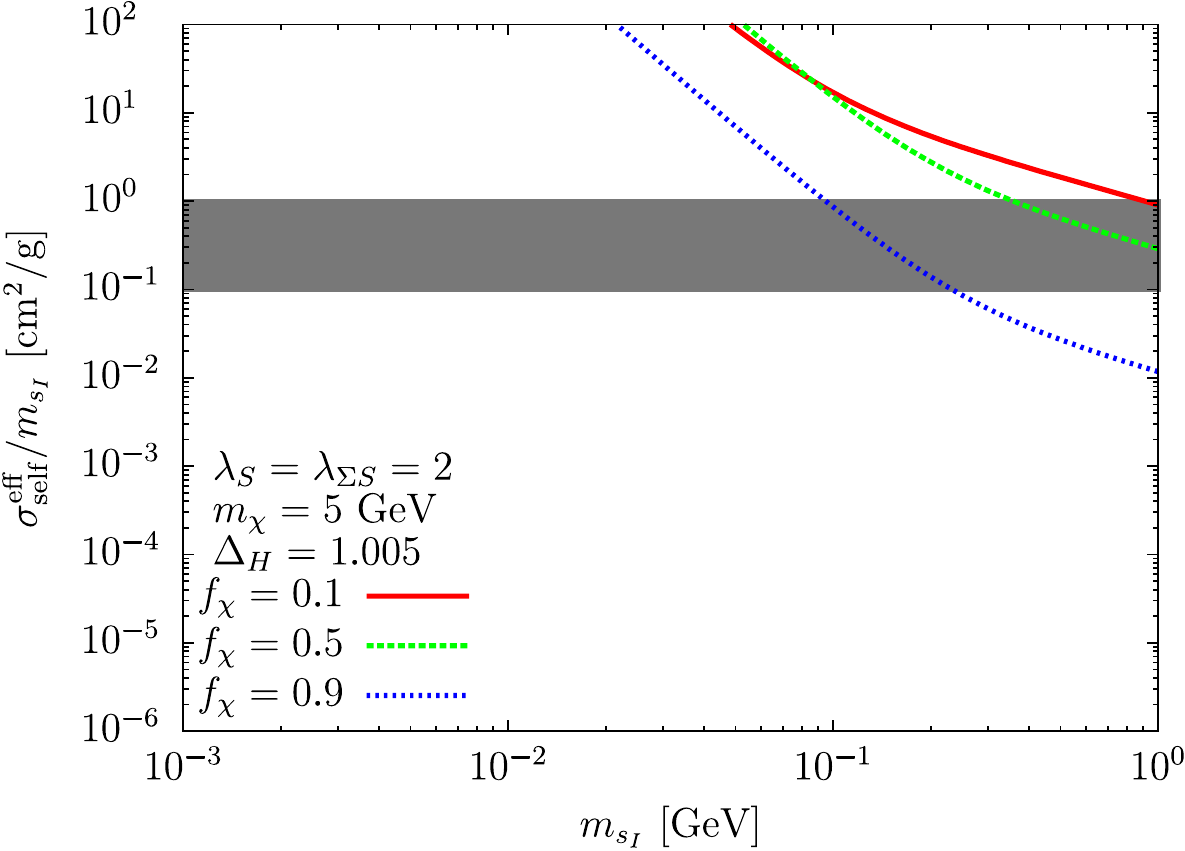}\\
\includegraphics[scale=0.65]{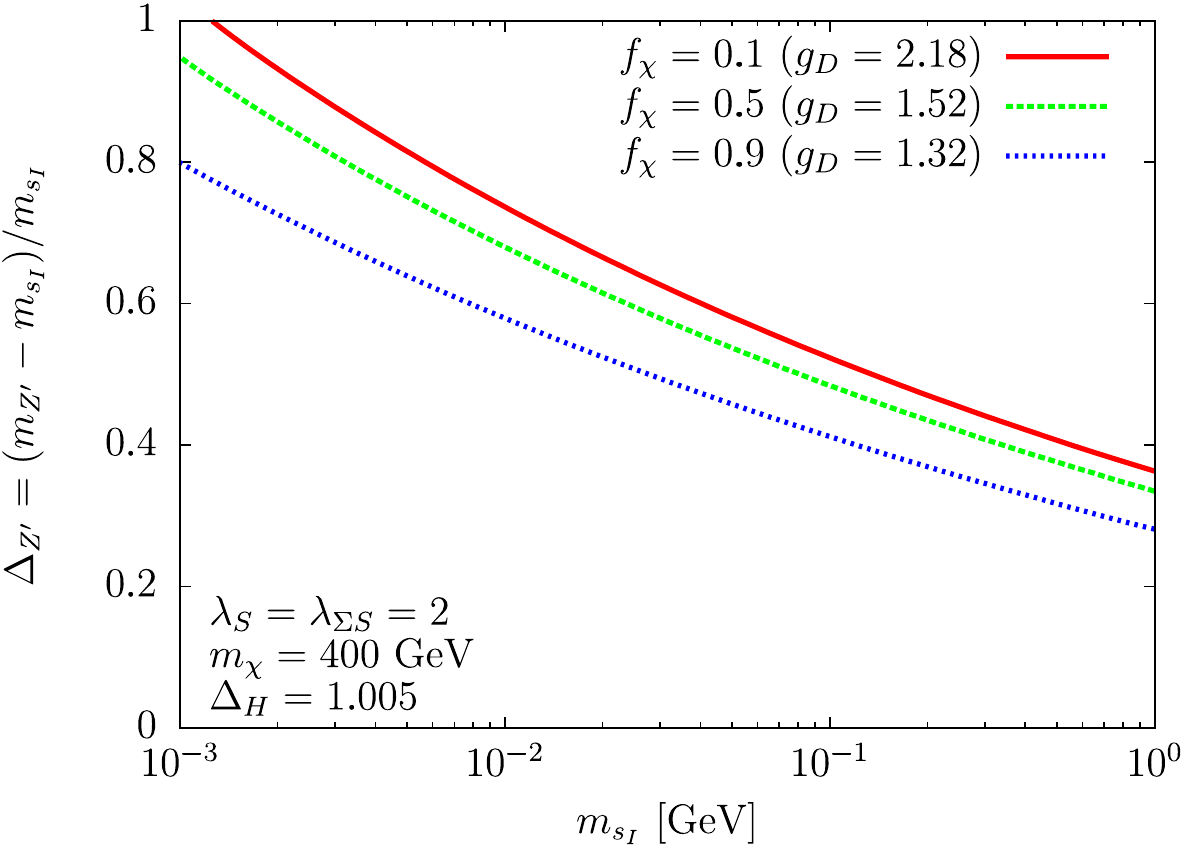}
\includegraphics[scale=0.65]{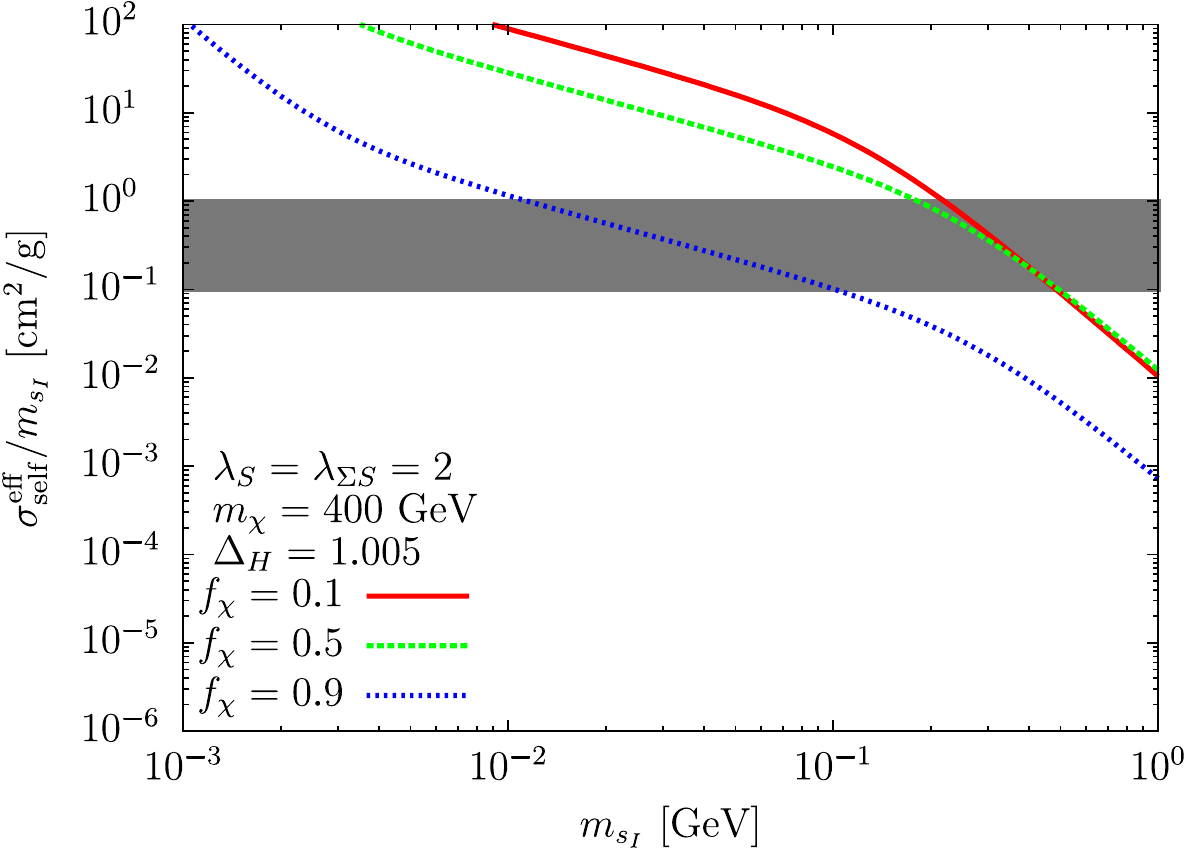}
\caption{Same plots as Fig.~\ref{fig:num1} but for $\Delta_H=1.005$.}
\label{fig:num2}
\end{center}
\end{figure}

The relic density and self-interacting cross section can numerically be computed. 
Fig.~\ref{fig:num0} shows the parameter space where the fraction 
$f_\chi\equiv
\Omega_\chi/\Omega_\mathrm{exp}$ gives
$0.1<f_\chi<0.9$ in the plane of ($m_\chi$, $g_D$) for $m_{Z^\prime}=100~\mathrm{MeV}$. 
The red and the green regions respectively give 
the results with and without the Sommerfeld effect.
One can see that the Sommerfeld effect becomes effective when
$m_\chi\gtrsim100~\mathrm{GeV}$, which makes  
the required value of $g_D$ smaller.\footnote{Implications of such a
Sommerfeld effect on BBN and CMB have been discussed in~\cite{Slatyer:2009yq,
Zavala:2009mi, Hisano:2011dc, Kawasaki:2015yya}.}

In the left plots in Fig.~\ref{fig:num1}, the contours generating 
the centre value of the observed relic density 
$\Omega_{s_I} h^2+\Omega_\chi h^2 = 0.12$ are shown in the plane of
$(m_{s_I}, \Delta_{Z^\prime})$ where the dark
matter mass $m_\chi$ is fixed to 
be $m_\chi=5~\mathrm{GeV}$ in the upper panel and $m_\chi=400~\mathrm{GeV}$
in the lower panel. 
The other relevant parameters are fixed to be
$\lambda_{S}=\lambda_{\Sigma S}=2$
and $\Delta_H=1.1$. 
The quartic couplings $\lambda_{\Phi\Sigma}$ and $\lambda_{\Phi S}$
should be small enough  to evade the constraint of the Higgs invisible
decay, and $\lambda_\Sigma$ should be taken such that
$\lambda_\Sigma\langle\Sigma\rangle\ll\lambda_{\Phi\Sigma}\langle\Phi\rangle$
for the reasonable mass matrix in the upper line of Eq.~(\ref{eq:higgs}). 
Each line of red, green and blue corresponds to
$f_\chi=0.1,0.5$ and $0.9$, respectively. 
When the $Z^\prime$ gauge boson mass is light enough
($m_{Z^\prime}\ll m_\chi$), the relic density of the heavier dark matter 
component $\chi$ is almost determined by 
two parameters,
$g_D$ and
$m_\chi$. Thus the hidden gauge coupling $g_D$ is also 
fixed for each colored line to get the fixed fraction $f_\chi$. 

From the plots, one can see that the $Z^\prime$ mass should be
$m_{Z^\prime}\lesssim2 m_{s_I}$ ($\Delta_{Z^\prime}\lesssim1$). 
In this parameter set, the forbidden channel $SS^\dag\to Z^\prime
Z^\prime$ is dominant in most of the parameter region since the
dark Higgs mass is fixed to be $m_H=2.1m_S$ ($\Delta_H=1.1$) which is
too heavy to induce the forbidden channel $SS^\dag\to HH$. 
In addition, one can see that the required value of $\Delta_{Z^\prime}$
becomes smaller for heavier $m_S$. 
This is because the thermally averaged cross section
$\langle\sigma{v}\rangle_{SS^\dag\to Z^\prime Z^\prime}$ given by
Eq.~(\ref{eq:forbidden}) is roughly scaled as
$\langle\sigma{v}\rangle_{SS^\dag\to Z^\prime Z^\prime}\sim
e^{-2\Delta_{Z^\prime} z_S}/m_S^2$, which should be almost constant to 
explain the observed relic density.
Therefore the decrease in the cross section due to heavier $m_S$ must be
compensated by the reduction of $\Delta_{Z^\prime}$. 

In the right plots in Fig.~\ref{fig:num1}, the effective
self-interacting cross section $\sigma_\mathrm{self}^\mathrm{eff}/m_{s_I}$
is shown as a function of $m_{s_I}$  
for $f_\chi=0.1,0.5$ and $0.9$. 
The relevant parameter set is the same with the left plots. 
The self-interacting cross
section as large as 
$0.1~\mathrm{cm^2/g}\leq\sigma_\mathrm{self}^\mathrm{eff}/m_{s_I}\leq1~\mathrm{cm^2/g}$
can be obtained around $m_{s_I}\sim \mathcal{O}(100)~\mathrm{MeV}$ for
$m_\chi=5~\mathrm{GeV}$ 
and $m_{s_I}\sim 3-80~\mathrm{MeV}$ for
$m_\chi=400~\mathrm{GeV}$.
As can be seen in the left plots in Fig.~\ref{fig:num1}, 
the $Z^\prime$ mass can be $\mathcal{O}(1-100)~\mathrm{MeV}$ 
to obtain the relic density consistent with the observed value.

The numerical results with $\Delta_H=1.005$ are shown in
Fig.~\ref{fig:num2}. 
Taking $\Delta_H=1.005$ means that the mass of the dark Higgs boson is
close to the resonance $2m_{s_I}\approx m_H$. 
Thus the self-interacting cross section given by Eq.~(\ref{eq:self_res})
can be enhanced as one can see from Fig.~\ref{fig:num1} and
Fig.~\ref{fig:num2}. 
We stress that the parameter region which can reproduce the appropriate value of
the effective self-interacting cross section is shifted to heavier
$m_{s_I}$. As a result, it is possible to take 
$m_{s_I},m_{Z^\prime},m_H=\mathcal{O}(100)~\mathrm{MeV}$ 
even for $m_\chi=400~\mathrm{GeV}$.

\section{Detection Properties}
\label{sec:4}
\subsection{Indirect Detection}
\begin{figure}[t]
\begin{center}
\includegraphics[scale=0.65]{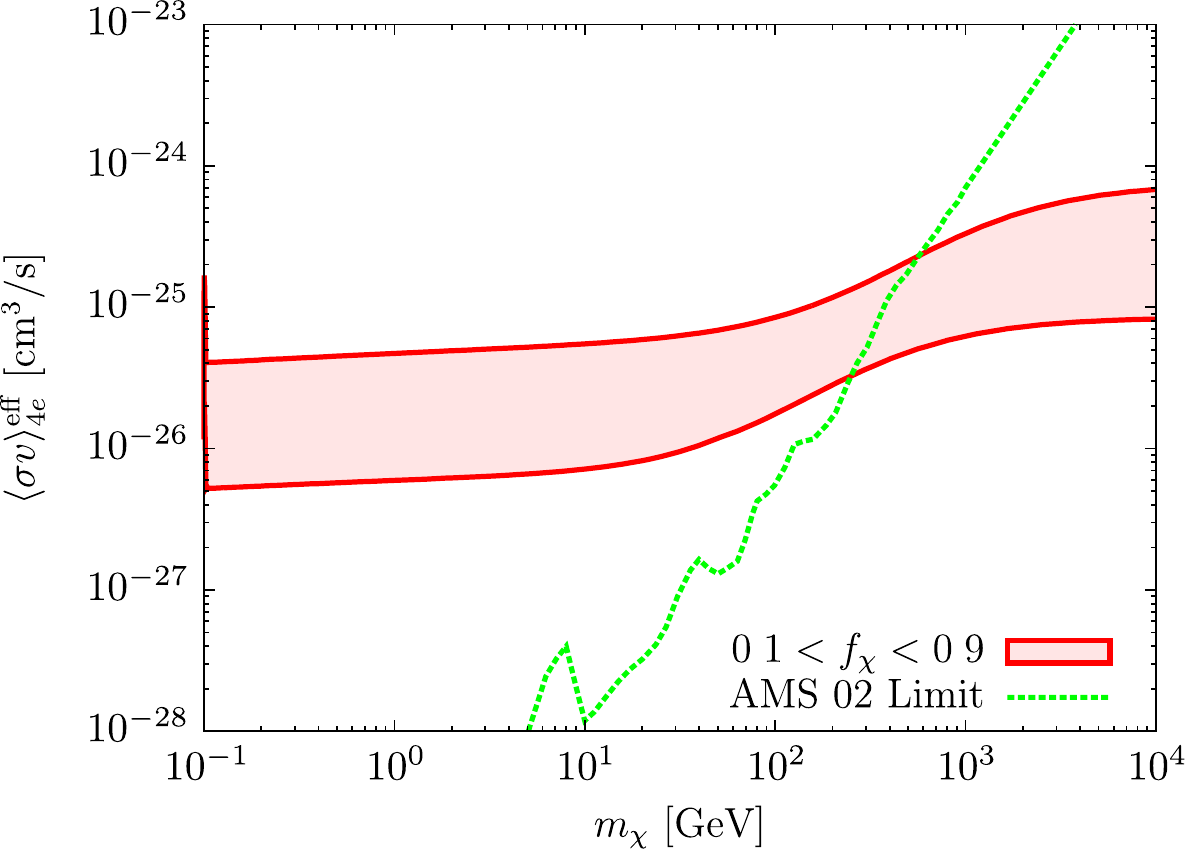}
\caption{AMS-02 constraint where the upper region of the green line is
 excluded. }
\label{fig:indirect}
\end{center}
\end{figure}

In this model, the energetic $e^\pm$ are produced through the annihilation 
of the heavier dark matter particle
$\chi\chi^\dag\to Z^\prime Z^\prime\to e^+e^-e^+e^-$ because the main decay mode of
the $Z^\prime$ gauge boson is $Z^\prime\to e^+e^-$.\footnote{Depending on
the $Z^\prime$ mass, the other decay channels are also possible such as
$\mu^+\mu^-$ and $\pi^+\pi^-$.} 
Furthermore, the annihilation cross section for this channel is enhanced
by the Sommerfeld effect of the light mediator 
$Z^\prime$. Thus, as can be seen in Eq.~(\ref{eq:sv2}),
 the parameters $m_\chi$ and $g_D$ relevant to the
annihilation cross section of $\chi$
are strongly
constrained by the AMS-02 positron observation~\cite{Aguilar:2013qda}. 
The constraints on the annihilation cross sections for the specific
channels such as $e^+e^-$, $\mu^+\mu^-$,
$\tau^+\tau^-$, $b\overline{b}$, $W^+W^-$ are given in
Ref.~\cite{Ibarra:2013zia}, and 
the constraint for the final state $e^+e^-$ is especially
strong.
In order to apply this bound for our case conservatively, we define the effective cross
section into $e^+e^-e^+e^-$ by
\begin{equation}
\langle\sigma{v}\rangle_{4e}^\mathrm{eff}\equiv
2\left(\frac{\Omega_\chi}{\Omega_\mathrm{exp}}\right)^2
\langle\sigma{v}\rangle_{\chi\chi^\dag \to Z^\prime Z^\prime},
\end{equation}
where the factor $2$ comes from two pairs of $e^+e^-$ generated for
each annihilation.\footnote{Precisely, since the energy of the produced
positrons and electrons is different for $e^+e^-$ and
$e^+e^-e^+e^-$ final states, the translation of the bound
for $e^+e^-$ into $e^+e^-e^+e^-$ discussed here is
not exactly true. Thus the obtained bound for $e^+e^-e^+e^-$
should be regarded as a conservative bound.} 
We impose the constraint that this effective cross section should be
smaller than the AMS-02 bound for the channel $e^+e^-$~\cite{Ibarra:2013zia}. 

The bound for $e^+e^-$ obtained by assuming the
Einasto profile and the MED propagation model in
Ref.~\cite{Ibarra:2013zia} is translated into 
$e^+e^-e^+e^-$ as shown in
Fig.~\ref{fig:indirect}. 
One should note that the upper bounds include some uncertainties such as
the dark matter density profiles and the diffusion models. 
From Fig.~\ref{fig:indirect}, 
one can see that $m_\chi\gtrsim800~\mathrm{GeV}$ is required to evade
the AMS-02 constraint for $f_\chi=0.9$, and
$m_\chi\gtrsim300~\mathrm{GeV}$ for $f_\chi=0.1$. 
Thus the mass of the dark matter component $\chi$ should be typically larger than
the electroweak scale in order to satisfy $0.1<f_\chi<0.9$. 

Due to this constraint, the numerical computations for
$m_{\chi}=5~\mathrm{GeV}$ in Fig.~\ref{fig:num1} and \ref{fig:num2} are
already excluded. For $m_\chi=400~\mathrm{GeV}$, only the blue line for
$f_\chi=0.9$ would be excluded. If a heavier $m_\chi$ is taken, 
$f_\chi$ can be larger, however a larger gauge coupling $g_D$ is
required to reproduce the correct relic density, which should be perturbative.

\subsection{Direct Detection}

\begin{figure}[t]
\begin{center}
\includegraphics[scale=0.75]{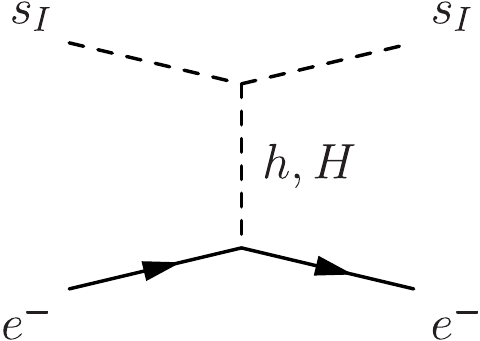}
\quad
\includegraphics[scale=0.75]{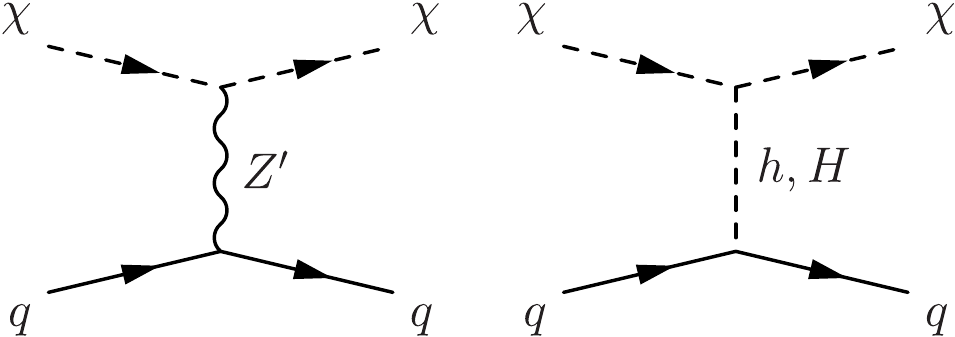}
\caption{Diagrams for direct detection of dark matter $s_I$ (left) and
 $\chi$ (centre and right).}
\label{fig:dd}
\end{center}
\end{figure}

Since the mass scale of the lighter dark matter particle $s_I$ in this
model is a few MeV to several hundred MeV, 
it would be difficult to detect it via direct detection with nuclei. 
However there is a chance to explore via direct detection with electron
as shown in the left diagram in Fig.~\ref{fig:dd}
where the elastic scattering of the lighter dark matter particle
$s_I$ with electron occurs via the Higgs couplings. 
The spin
independent cross section is computed as 
\begin{equation}
\sigma_\mathrm{DD}^S=
\frac{m_e^4}{4\pi\left(m_e+m_{s_I}\right)^2}
\left(\frac{\mu_{Hs_Is_I}\sin\alpha}{m_H^2\langle\Phi\rangle}
-\frac{\mu_{hs_Is_I}\cos\alpha}{m_h^2\langle\Phi\rangle}\right)^2,
\label{eq:dd1}
\end{equation}
where $\mu_{hs_Is_I}~(=\mu_{hSS^\dag})$ and
$\mu_{Hs_Is_I}~(=\mu_{HSS^\dag})$ are given by Eq.~(\ref{eq:SSh}) and
(\ref{eq:HSS}), respectively. 
The cross section is very suppressed by the electron Yukawa coupling
(the electron mass).
The scattering event with electron can be searched by direct detection
experiments with current technology~\cite{Essig:2011nj, Hochberg:2015pha}. 
The current strongest upper bound for the elastic cross section is given by
XENON10
as $\sigma_\mathrm{DD}\lesssim10^{-38}~\mathrm{cm}^2$~\cite{Essig:2012yx} 
at the dark matter mass of around $100~\mathrm{MeV}$. 
On the other hand, the typical order of the cross section given by
Eq.~(\ref{eq:dd1}) is 
$\sigma_\mathrm{DD}^S\sim10^{-45}~\mathrm{cm}^2$ for
$m_{s_I}\sim m_H\sim100~\mathrm{MeV}$ and
$\mu_{hs_Is_I}\sim\mu_{Hs_Is_I}\sim1~\mathrm{GeV}$, thus this constraint
is easily satisfied. 
The experimental sensitivity can reach
to the order of $\sigma_\mathrm{DD}\sim10^{-43}~\mathrm{cm}^2$ with future
experiments~\cite{Essig:2011nj}. In particular, high sensitivity is
achieved for the experiments using Germanium since the ionization
threshold of Germanium is lower than the other elements such as Xenon
and Argon. 

\begin{figure}[t]
\begin{center}
\includegraphics[scale=1]{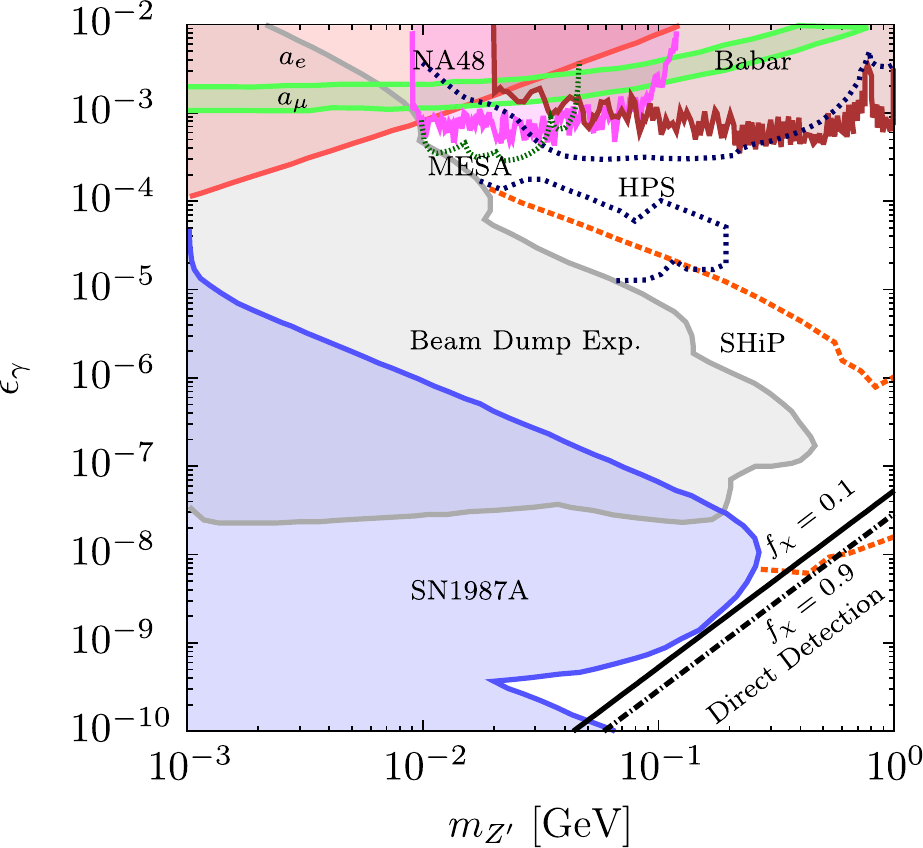}
\caption{Current constraint on the kinetic mixing $\epsilon_\gamma$ and
 future sensitivities including beam dump
 experiments~\cite{Essig:2013lka},
 HPS~\cite{Essig:2013lka}, 
 SN1987A~\cite{Kazanas:2014mca}, NA48/2~\cite{Batley:2015lha},
 Babar~\cite{Lees:2014xha}, MESA~\cite{Beranek:2013yqa},
 SHiP~\cite{Alekhin:2015byh}. The black lines 
 correspond to the upper bounds obtained by direct detection of the heavier dark
 matter particle $\chi$ when the $Z^\prime$ mediated contribution is
 dominant in Eq.~(\ref{eq:dd2}).}
\label{fig:eps_mz}
\end{center}
\end{figure}

For the heavier dark matter particle $\chi$, 
since the mass should be above the electroweak scale as derived from the
indirect detection constraint, the most stringent
bound on the relevant parameters is given by the elastic scatting
with a proton as shown in the centre and right diagrams in Fig.~\ref{fig:dd}. 
The elastic scattering cross section for $\chi$ with a
proton is computed as
\begin{equation}
\sigma_{\mathrm{DD}}^\chi=
\frac{m_p^2}{16\pi\left(m_p+m_\chi\right)^2}
\left[\frac{g_De\epsilon_\gamma m_\chi}{m_{Z^\prime}^2}
+\left(\sum_q f_q^p\right)\left(
\frac{\mu_{H\chi\chi^\dag}\sin\alpha}{m_H^2}
-\frac{\mu_{h\chi\chi^\dag}\cos\alpha}{m_h^2}\right)\frac{2m_p}
{\langle\Phi\rangle}\right]^2,
\label{eq:dd2}
\end{equation}
where $m_p$ is the proton mass $m_p=938~\mathrm{MeV}$, the
coefficient $f_q^p$ is given in Ref.~\cite{Cline:2013gha} and
\begin{equation}
\mu_{H\chi\chi^\dag}=
\sqrt{2}\left(-\lambda_{\Phi\chi}\langle\Phi\rangle\sin\alpha
+\lambda_{\Sigma\chi}\langle\Sigma\rangle\cos\alpha\right).
\end{equation}
The first and second terms in Eq.~(\ref{eq:dd2}) correspond to the
centre and right diagrams in Fig.~\ref{fig:dd}, respectively. 
This spin independent cross section for $\chi$ is strongly constrained because the mediators
$Z^\prime$ and $H$ are much lighter than the dark matter mass $m_\chi$. 
The effective spin-independent cross section defined by
$\sigma_\mathrm{DD}^\chi (\Omega_\chi/\Omega_\mathrm{exp})$
should be
compared with the experimental limits since 
the limits are derived for one-component dark matter.
Under our assumptions 
$\lambda_{\Phi\chi},\lambda_{\Sigma\chi}\ll1$,
we focus on the case that the 
$Z^\prime$ gauge boson contribution is dominant (the first term in Eq.~(\ref{eq:dd2})). 
In this case, the upper bound on the kinetic mixing $\epsilon_\gamma$ is obtained
depending on the fraction $f_\chi$ of the total relic density as
shown by the black lines in Fig.~\ref{fig:eps_mz}. 
The upper region of the black lines are excluded by the direct detection experiments.
The colored regions, except for the green region, have already been
excluded by the experiments 
(beam dump experiments~\cite{Essig:2013lka}, HPS~\cite{Essig:2013lka}, 
 SN1987A~\cite{Kazanas:2014mca}, NA48/2~\cite{Batley:2015lha},
 Babar~\cite{Lees:2014xha}, MESA~\cite{Beranek:2013yqa}, SHiP~\cite{Alekhin:2015byh}).
The kinetic mixing gives a new contribution to anomalous magnetic moment
of a charged lepton 
and the green region in Fig.~\ref{fig:eps_mz} can account for the
deviation between the experiment and the SM prediction
for muon anomalous magnetic moment~\cite{Lee:2014tba}. 
The required order of the kinetic mixing is roughly
$\epsilon_\gamma\sim10^{-3}$, while $\epsilon_\gamma\lesssim10^{-7}$ or
$10^{-8}$ is needed in our scenario of multi-component dark matter
because of the strong direct detection bound for the heavier dark matter 
particle $\chi$. 
In addition, the region of $\epsilon_\gamma\sim10^{-3}$ has already been
excluded by the other experiments. 

In order to be consistent with getting the large self-interacting
cross section
for solving the small scale problems,
$0.1~\mathrm{cm^2/g}\leq\sigma_\mathrm{self}^\mathrm{eff}/m_{s_I}\leq1~\mathrm{cm^2/g}$
, we found in Section~\ref{sec:3} that 
the mass of $Z^\prime$ gauge boson is $m_{Z^\prime}\lesssim1~\mathrm{GeV}$. 
Fig.~\ref{fig:eps_mz} shows that the kinetic mixing $\epsilon_\gamma$ should be
$\epsilon_\gamma\sim10^{-10}-10^{-9}$ for
$m_{Z^\prime}\sim\mathcal{O}(100)~\mathrm{MeV}$ 
and $\epsilon_\gamma\sim10^{-9}-10^{-8}$ for
$m_{Z^\prime}\gtrsim\mathcal{O}(100)~\mathrm{MeV}$ with $0.1<f_\chi<0.9$. 
The region of $m_{Z^\prime}\lesssim50~\mathrm{MeV}$ is excluded by the
constraint of SN1987A~\cite{Kazanas:2014mca}.
Even for such a small kinetic mixing, some parameter space 
with $m_{Z^\prime}$ of several hundred MeV
can be tested by the SHiP (Search for Hidden
Particles) experiment which is a newly proposed proton beam on target (tungsten)
experiment~\cite{Gorbunov:2014wqa, Alekhin:2015byh}.
One more point is that a cancellation between the two
different contributions mediated by the $Z^\prime$ gauge boson and the
Higgs bosons may be possible (see Eq.~(\ref{eq:dd2})). 
In this case, the strong upper bound for
the kinetic mixing would partially be relaxed.
Since the magnitude of the kinetic mixing
$\epsilon_\gamma$ is very small, the dark matter particles cannot be in
kinetic equilibrium with the SM particles via the kinetic mixing. 
Instead of that the thermalization with the SM particles is realized
by the couplings in the scalar potential.


\section{Summary and Conclusions}
\label{sec:5}
We have considered the model extended with the hidden $U(1)_D$ gauge
symmetry. 
Due to the remnant $\mathbb{Z}_4$ symmetry, the hidden scalar $\chi$ can
be a dark matter candidate. 
In addition, since the decay of the light scalar $s_I$ is kinematically
forbidden, $s_I$ can be a second dark matter component. 
We have discussed the phenomenology of 
the two dark matter components.
The relic density of the lighter dark matter $s_I$ can be determined by
the forbidden channels 
into the $Z^\prime$ gauge boson and the extra dark Higgs boson $H$ 
while that of the heavier state $\chi$ is dictated
by the
normal annihilation channel into the $Z^\prime$ gauge boson. 
In this framework, the mass scale of the lighter dark matter particle
$s_I$ is typically $1~\mathrm{MeV}\lesssim
m_{s_I}\lesssim\mathcal{O}(100)~\mathrm{MeV}$ and the couplings can be
larger than the typical WIMP or WIMPless dark matter because the relic
density of $s_I$ is produced 
by forbidden channels, thus the lighter dark matter particle $s_I$ can generate a
large self-interacting cross section to solve the small scale problems of $\Lambda$CDM model. 

We have also taken into account the constraints of indirect detection
and direct detection of dark matter. 
From the constraint of indirect detection, the mass of the heavier dark
matter state $\chi$ should be larger than electroweak scale. 
The constraint of direct detection for the heavier dark
matter $\chi$ is very strong since the elastic scattering with a nucleon
can be induced by the light $Z^\prime$ gauge boson. To avoid this
constraint, the kinetic mixing $\epsilon_\gamma$ and the scalar couplings
between the heavier dark matter particle $\chi$ and the Higgs bosons should be very small. 
Nevertheless, if the $Z^\prime$ gauge boson is
$m_{Z^\prime}\gtrsim200~\mathrm{MeV}$ and
$\epsilon_\gamma\gtrsim10^{-8}$, the model can be tested by the SHiP
future experiment. 


\section*{Acknowledgments}
The work of M.~A. is supported in part by the Japan Society for the
Promotion of Sciences (JSPS) Grant-in-Aid for Scientific Research (Grant
No. 25400250 and No. 16H00864). 
T.~T. acknowledges support from P2IO Excellence Laboratory (LABEX) and
JSPS Fellowships for Research Abroad. 
T.~T. would like to thank Yann Mambrini for fruitful discussion. 


\end{document}